\documentclass[fleqn,usenatbib,useAMS]{mnras}

\usepackage{mathptmx}
\usepackage[T1]{fontenc}
\usepackage{ae,aecompl}

\usepackage{graphicx}	% Including figure files
\usepackage{amsmath}	% Advanced maths commands
\usepackage{amssymb}	% Extra maths symbols
\usepackage{bm}		% Bold maths symbols, including upright Greek
\usepackage{pdflscape}	% Landscape pages
\usepackage{color}
\usepackage{exscale}
\usepackage{natbib}
\usepackage{rotating}
\usepackage{rotate}
\usepackage{afterpage}
\usepackage{booktabs,threeparttable}
\usepackage{relsize}
\usepackage{lscape}
\usepackage{tabularx}
\usepackage{longtable}
\usepackage{ltxtable}
\usepackage{txfonts}
\usepackage{natbib}
\usepackage{bbm}
\usepackage{afterpage}
\usepackage{widetext}

\def\apj{ApJ}%
\def\apjl{ApJ}%
\def\apjs{ApJS}%
\def\ao{Appl.~Opt.}%
\def\aap{A\&A}%
\def\mnras{MNRAS}%
\def\prd{Phys.~Rev.~D}%
\def\jcap{JCAP}%
\def\physrep{Phys.~Rep.}%
\newcommand{\tbf}{\textbf}
\newcommand{\ti}{\textit}

\newcommand{\bea}{\begin{eqnarray}}
\newcommand{\be}{\begin{equation}}
\newcommand{\ben}{\begin{enumerate}}
\newcommand{\bi}{\begin{itemize}}
\newcommand{\eea}{\end{eqnarray}}
\newcommand{\ee}{\end{equation}}
\newcommand{\ei}{\end{itemize}}
\newcommand{\een}{\end{enumerate}}

\newcommand{\matC}{\mathbf C}

\newcommand{\ensav}[1]{\left\langle #1 \right\rangle}

\newcommand{\like}{L}
\newcommand{\prob}{P}
\newcommand{\probr}{P_r}

\newcommand{\pco}{\vek p_\mr{c}}
\newcommand{\pnu}{\vek p_\mr{n}}

\newcommand{\D}{\vek D}

\newcommand{\M}{\vek M}

\newcommand{\om}{\Omega_\mr m}
\newcommand{\omb}{\Omega_\mr b}
\newcommand{\sig}{\sigma_8}
\newcommand{\ns}{n_s}
\newcommand{\w}{w_0}
\newcommand{\wa}{w_a}

\renewcommand{\d}{{\rm d}}

\newcommand{\mr}{\mathrm}

\newcommand{\vek}{\mathbf}

\usepackage{titlesec}
\titlespacing*{\paragraph}
{0pt}{0.2cm}{0.3cm}

\title[\textsc{CosmoLike}]{\textsc{CosmoLike} - Cosmological Likelihood Analyses for Photometric Galaxy Surveys}

\author[Krause \& Eifler]{Elisabeth Krause$^{1}$\thanks{E-mail: aekrause@stanford.edu}, Tim Eifler$^{2,3}$\thanks{E-mail: tim.eifler@jpl.nasa.gov} \\
$^{1}$ Kavli Institute for Particle Cosmology and Astrophysics, Stanford University, Stanford, CA 94305, USA\\
$^{2}$ Jet Propulsion Laboratory, California Institute of Technology, Pasadena, CA 91109, USA\\
$^{3}$  Department of Physics, California Institute of Technology, Pasadena, CA 91125, USA }
\begin{document}

\date{Accepted . Received ; in original form }

\pagerange{\pageref{firstpage}--\pageref{lastpage}} \pubyear{2014}

\maketitle

\label{firstpage}

\begin{abstract}
We explore strategies to extract cosmological constraints from a joint analysis of cosmic shear, galaxy-galaxy lensing, galaxy clustering, cluster number counts, and cluster weak lensing. We utilize the \textsc{CosmoLike} software to simulate results from an LSST like data set, specifically, we 1) compare individual and joint analyses of the different probes, 2) vary the selection criteria for lens and source galaxies, 3) investigate the impact of blending, 4) investigate the impact of the assumed cosmological model in multi-probe covariances, 6) quantify information content as a function of scales, and 7) explore the impact of intrinsic galaxy alignment in a multi-probe context. Our analyses account for all cross correlations within and across probes and include the higher-order (non-Gaussian) terms in the multi-probe covariance matrix. We simultaneously model cosmological parameters and a variety of systematics, e.g. uncertainties arising from shear and photo-z calibration, cluster mass-observable relation, galaxy intrinsic alignment, and galaxy bias (up to 54 parameters altogether).\\
We highlight two results: First, increasing the number density of source galaxies by $\sim$30\%, which corresponds to solving blending for LSST, only gains little information. Second, including small scales in clustering and galaxy-galaxy lensing, by utilizing HODs, can substantially boost cosmological constraining power.\\ 
The \textsc{CosmoLike} modules used to compute the results in this paper will be made publicly available at https://github.com/elikrause/CosmoLike\_Forecasts.
\end{abstract}


\renewcommand{\thefootnote}{\arabic{footnote}}
\setcounter{footnote}{0}

\section{Introduction}
The Universe is a fascinating physics laboratory with the unfortunate limitation that we cannot influence the settings of its experiments. We can only observe as many of the Universes' `features' as possible, and interconnect these observables to either falsify our physical models or to tighten corresponding constraints.

Ongoing photometric surveys, such as Kilo-Degree Survey (KiDS\footnote{http://www.astro-wise.org/projects/KIDS/}), Hyper Suprime Cam (HSC\footnote{http://www.naoj.org/Projects/HSC/HSCProject.html}), and Dark Energy Survey (DES\footnote{www.darkenergysurvey.org/}) will provide an order of magnitude increase in high-quality imaging data in the very near future. These data sets provide an exceptional opportunity to study the physics of the Universe (e.g., cosmic acceleration, neutrino mass, tests of General Relativity), with the prospect of even more and deeper data from next generation experiments, such as the Large Synoptic Survey Telescope (LSST\footnote{http://www.lsst.org/lsst}), Euclid\footnote{sci.esa.int/euclid/} and the Wide-Field Infrared Survey Telescope (WFIRST\footnote{http://wfirst.gsfc.nasa.gov/}).  

Multiple probes tracing the Large Scale Structure (LSS) of the Universe can be extracted from these photometric data sets, e.g. cosmic shear, galaxy-galaxy lensing, galaxy clustering, Baryon Acoustic Oscillations (BAO), galaxy cluster number counts, and galaxy cluster weak lensing. This variety enables powerful analysis strategies to advance our understanding of cosmology: First, inconsistencies among different probes can indicate new physics. Second, the joint analysis of consistent probes significantly tightens constraints on the evaluated cosmological model \citep[see e.g.,][]{wme13}. 

It is straightforward to compare/combine individual LSS probes with Supernovae 1a (SN1a) or Cosmic Microwave Background (CMB) measurements when assuming that corresponding information is independent \citep[e.g.,][]{kfh13,DES15WL}. Complications arise quickly when multiple LSS probes are included in the analysis, since these have non-negligible correlations in their signals as well as in their systematics. 

Multi-probe analyses have been suggested in the literature as the most promising way to constrain cosmology. For example, \citep{ber09} give a detailed description of a Fisher matrix analysis of galaxy clustering, cosmic shear, and galaxy-galaxy lensing. Similar analyses are presented in \cite{job10} and \cite{yos12}, where the latter considers number counts of galaxy clusters instead of cosmic shear. All three analyses use so-called Gaussian covariances, where Gaussian means that connected higher-order moments of the density field are not included in the covariance computation. However, covariance terms that arise from these higher-order moments can significantly impact error bars \citep{sht09}; these terms  were included in the analyses of \cite{eks14, tas14, pkd15} all of which vary in terms of the probes considered. 

In this paper we take simulated multi-probe analyses to a new level and present joint analyses of cosmic shear, galaxy-galaxy lensing, galaxy clustering, photometric BAO, galaxy cluster number counts, and galaxy cluster weak lensing. We model all cross-correlations among probes, use analytical non-Gaussian covariances in the simulated likelihood analysis, and include uncertainties from various systematics (photo-z and galaxy shape measurements, galaxy bias models, cluster-mass observable relation, and galaxy intrinsic alignments). Our metric to visualize the performance of different choices in the analysis is the $w_\mr p - \wa$ plane, where $z_\mr p$ denotes a pivot redshift at which both quantities de-correlate. 

Since we are only interested in relative performances of various analysis concepts \ti{we do not include any axis values in our plots}. It is important to keep in mind that absolute values characterizing survey performance depend on a variety of assumptions and that these values should not be over-interpreted. We will make corresponding \textsc{CosmoLike} modules publicly available at https://github.com/elikrause/CosmoLike\_Forecasts upon acceptance of the paper. This repository also contains plots with axis values for those who cannot resist.

%%%%%%%%%%%%%%%%%%%%%%%%%%%%%%%%%%%%%%%%%%%%%%%%%%
%%%%%%%%%%%%%%%%%%%%%%%%%%%%%%%%%%%%%%%%%%%%%%%%%%
\section{Ingredients of the analysis}
%%%%%%%%%%%%%%%%%%%%%%%%%%%%%%%%%%%%%%%%%%%%%%%%%%
%%%%%%%%%%%%%%%%%%%%%%%%%%%%%%%%%%%%%%%%%%%%%%%%%%
We first clarify terminology employed throughout the paper:
\paragraph*{Observables} Fields constructed from the survey catalogs, such as the convergence field $\kappa$, projected galaxy/cluster density, as well the associated density contrast fields $\delta_{\mathrm{g}},\delta_{\lambda_\alpha}$. 
\paragraph*{Probe} Combination of $n$ observables (including $n=1$). Example: $\kappa \kappa$, also termed cosmic shear.
\paragraph*{Summary statistic} Data compression of the probes, which hopefully preserves the information content. In this paper we employ number counts ($n =1$), and 2-point functions. Other possibilities are higher-order correlation functions, Minkowski functionals \citep[e.g.,][]{klw12}, and other classifications of the density field \citep[e.g.,][]{ljw15}.
\paragraph*{Data/model vector} A vector consisting of data points that are measured/modeled summary statistics. 

%%%%%%%%%%%%%%%%%%%%%%%%%%%%%%%%%%%%%%%%%%%%%%%%%%
\subsection{Summary statistics}
\label{sec:sumstats}
%%%%%%%%%%%%%%%%%%%%%%%%%%%%%%%%%%%%%%%%%%%%%%%%%%
We compute summary statistics from three types of objects that can be extracted from a photometric survey: 

\paragraph*{Lens galaxies,} characterized through positions and redshift estimates for a specific sample definition.
\paragraph*{Source galaxies,} characterized through positions, redshift and shape estimates for the source galaxy sample.
\paragraph*{Galaxy clusters,} characterized through positions, redshift and optical richness estimates, serving as an observable proxy for cluster mass, for galaxy clusters selected from the galaxy catalog.

Each catalog is split into (photometric) redshift bins, which we denote with a lower case Roman superscript, and we furthermore divide the cluster catalog into bins of optical richness, which denote with a lower case Greek subscript.
In this analysis, we focus on number counts and 2-point statistics constructed from the three catalogs introduced above. In particular, we choose to work with angular (cross) power spectra, as opposed to angular correlation functions, for the 2-point statistics for computational speed. Throughout this analysis we consider only cosmological models without spatial curvature.
\subsubsection{Cluster Number Counts}
The expected cluster count in richness bin $\alpha$, with $\lambda_{\alpha,\mathrm{min}}< \lambda < \lambda_{\alpha,\mathrm{max}}$, and redshift bin $i$ with $z^i_{\lambda,\mathrm{min}}<z<z^i_{\lambda,\mathrm{max}}$ is given by
\begin{equation}
\mathcal{N}^i(\lambda_\alpha) = \Omega_{\mathrm s}\int_{z^i_{\lambda,\mathrm{min}}}^{z^i_{\lambda,\mathrm{max}}}dz \frac{d^2V}{dz d\Omega} \int dM \frac{dn}{dM}\int_{\lambda_{\alpha,\mathrm{min}}}^{\lambda_{\alpha,\mathrm{max}}} d\lambda\, p(M | \lambda,z)\,,
\label{eq:N}
\end{equation}
where $d^2V/dzd\Omega$ is the comoving volume element, $dn/dM$ the halo mass function in comoving units (for which we omitted the redshift dependence), $p(\lambda|M,z)$ is the probability distribution function that a dark halo of mass $M$ at redshift $z$ hosts a cluster with richness $\lambda$. 
Throughout this paper we define halo properties using the over density  $\Delta = 200 \bar{\rho}$, with $\bar{\rho}$ the mean matter density, and employ the \citet{tkk08,trk10} fitting function for the halo mass function.
We model the mean mass-observable relation $\bar{M}(\lambda)$ as a power-law in richness and redshift,
\be
\ln\left[\frac{\bar{M}(\lambda)}{M_\odot/h}\right] = C_\lambda+ a_{\lambda}\ln\left[\frac{\lambda}{60}\right] + b_{\lambda}\ln\left[1+z\right] 
\label{eq:Mlambda}
\ee
with normalization $c_{\lambda}$, slope $a_{\lambda}$ and redshift dependence $b_{\lambda}$,
and further assume a log-normal distribution with scatter $\sigma_{\mathrm{ln}M|\lambda}$:
\be
p(M| \lambda,z) = \frac{1}{M \sqrt{2\pi}\sigma_{\mathrm{ln}M|\lambda}} \exp \left[-\frac{(\ln[M]-\ln[\bar{M}(\lambda)])^2}{2\sigma_{\mathrm{ln}M|\lambda}}\right]\,.
\ee
\subsubsection{Power Spectra}
In this section we summarize the computation of angular (cross) power spectra for the different probes;  a more detailed derivation can be found in \citet{huj04}. We use capital Roman subscripts to denote observables, 
$A,B\in \left\{ \kappa,\delta_{\mathrm{g}},\delta_{\lambda_\alpha}\right\}$, 
where $\kappa$ references lensing, $\delta_{\mathrm{g}}$  the density contrast of (lens) galaxies, and $\delta_{\lambda_\alpha}$ the density contrast of galaxy clusters in richness bin $\alpha$.

We calculate the angular power spectrum between redshift bin $i$ of observable $A$ and redshift bin $j$ of observables $B$ at projected Fourier mode $l$, $C_{AB}^{ij}(l)$, using the Limber and flat sky approximations:
\begin{equation}
\label{eq:projected}
C_{AB}^{ij}(l) = \int d\chi \frac{q_A^i(\chi)q_B^j(\chi)}{\chi^2}P_{AB}(l/\chi,z(\chi)),
\end{equation}
where $\chi$ is the comoving distance, $q_A^i(\chi)$ are weight functions of the different observables 
given in Eqs.~(\ref{eq:qg}-\ref{eq:qkappa}), and $P_{AB}(k,z)$ the three dimensional, probe-specific power spectra detailed below. 
The weight function for the projected galaxy density in redshift bin $i$,$q_{\delta_{\mathrm{g}}}^i(\chi)$, is given the normalized comoving distance probability of galaxies in this redshift bin
\begin{equation}
\label{eq:qg}
q_{\delta_{\mathrm{g}}}^i(\chi) =\frac{n_{\mathrm{lens}}^i(z(\chi)) }{\bar{n}_{\mathrm{lens}}^i}\frac{dz}{d\chi}\,,
\end{equation}
with $n_{\mathrm{lens}}^i(z)$ the redshift distribution of galaxies in (photometric) galaxy redshift bin $i$ (c.f. Eq.~{\ref{eq:photoz}}), and $\bar{n}_{\mathrm{lens}}^i$ the angular number densities of galaxies in this redshift bin (c.f. Eq.~{\ref{eq:nbar}}). For the purpose of the forecasts presented here, we neglect variations of the cluster selection function within redshift bins, as well as uncertainties in the cluster redshift estimate; thus the weight function for the projected cluster density is given by
\begin{equation}
\label{eq:qlambda}
q_{\delta_{\lambda_\alpha}}^i(\chi) =  \Theta\left(z(\chi)-z^i_{\lambda,\mathrm{min}}\right)\Theta\left(z^i_{\lambda,\mathrm{max}}-z(\chi)\right)\frac{dV}{d\chi d\Omega}\,,
\end{equation}
with $\Theta(x)$ the Heaviside step function.
For the convergence field, the weight function $q_\kappa^{i}(\chi)$ is the lens efficiency, 
\begin{equation}
\label{eq:qkappa}
q_\kappa^{i}(\chi) = \frac{3 H_0^2 \Omega_m }{2 \mathrm{c}^2}\frac{\chi}{a(\chi)}\int_\chi^{\chi_{\mr h}} \mr d \chi' \frac{n_{\mathrm{source}}^{i} (z(\chi')) dz/d\chi'}{\bar{n}_{\mathrm{source}}^{i}} \frac{\chi'-\chi}{\chi'} \,.
\end{equation}
the lens efficiency, with $n_{\mathrm{source}}^{i} (z)$ the the redshift distribution of source galaxies in (photometric) source redshift bin $i$ (Eq.~{\ref{eq:photoz}}), $\bar{n}_{\mathrm{source}}^i$ the angular number densities of source galaxies in this redshift bin (Eq.~{\ref{eq:nbar}}), and $a(\chi)$ the scale factor.

We model the three-dimensional power spectra $P_{AB}(k,z)$ based on the non-linear matter power spectrum $P_{\mathrm{mm}}(k,z) =P_{\mathrm{NL}}(k,z)$, for we use the \citet{tsn12} fitting formula, or the halo model \citep{bew02,sel00,caw10} if one of the observables is not a linear projection of the matter density contrast. Noting that $P_{AB} = P_{BA}$, we describe the different cases in Eqs. (\ref{eq:Pkappa}-\ref{eq:Pcluster}).
For $A = \kappa$, this is trivial,
\begin{equation}
\label{eq:Pkappa}
P_{\kappa B}(k,z) = P_{mB}(k,z) \,.
\end{equation}
For the baseline data vector, we only consider the large-scale galaxy distribution, and assume that the galaxy density contrast on these scales can be approximated as the non-linear matter density contrast times an effective galaxy bias parameter $b_g(z)$  (but c.f. Sect.~\ref{sec:Rvary} for extensions),
\begin{equation}
\label{eq:Pg} P_{\delta_{\mathrm{g}} B}(k,z)  = b_g(z) P_{mB}(k,z) \,,
\end{equation}
and we model the redshift dependence of $b_g(z)$ within a redshift bin assuming passive evolution \citep{Fry96}. 

Within the halo model, the cross power spectrum between cluster centers and matter density contrast can be written as the usual sum of two- and one-halo term,
\begin{eqnarray}
\label{eq:Pcluster}
\nonumber P_{\delta_{\lambda_\alpha} m}(k,z) &\approx& b_{\lambda_\alpha}(z) P_{\mathrm{lin}}(k,z)\\
&+& \frac{ \int dM  \frac{dn}{dM}\frac{M}{\bar{\rho}} \tilde{u}_{\mathrm{m}}(k,M) \int_{\lambda_{\alpha,\mathrm{min}}}^{\lambda_{\alpha,\mathrm{max}}} d\lambda \,p(M | \lambda,z)}{\int dM \frac{dn}{dM}  \int_{\lambda_{\alpha,\mathrm{min}}}^{\lambda_{\alpha,\mathrm{max}}} d\lambda\, p(M | \lambda,z)},
\end{eqnarray}
with $P_{\mathrm{lin}}(k,z)$ the linear matter power spectrum. The mean linear bias of clusters in richness bin $\alpha$ reads 
\begin{equation}
\label{eq:blambda}
b_{\lambda_\alpha}(z) =  \frac{ \int dM  \frac{dn}{dM} b_{\mathrm{h}}(M)\int_{\lambda_{\alpha,\mathrm{min}}}^{\lambda_{\alpha,\mathrm{max}}} d\lambda \,p(M | \lambda,z)}{\int dM   \frac{dn}{dM}\int_{\lambda_{\alpha,\mathrm{min}}}^{\lambda_{\alpha,\mathrm{max}}} d\lambda p(M | \lambda,z)}\,,
\end{equation}
where $b_{\mathrm{h}}(M)$ the halo bias relation, for which we use the fitting function of \citet{trk10}. The Fourier transform of the radial matter density profile within a halo of mass $M$, $\tilde{u}_{\mathrm{m}}(k,M)$, is modeled assuming \citet{NFW}(NFW) profiles with the \citet{bhh11} mass-concentration relation $c(M,z)$,
\begin{eqnarray}
 \nonumber \tilde{u}_{\mathrm{m}}(k,M) &= \left[\ln(1+c(M)) -\frac{c(M)}{1+c}\right]^{-1}\bigg\{\sin(x)\left[\mathrm{Si}([1+c(M)]x-\mathrm{Si}(x)\right]\\
 &+\cos(x)\left[\mathrm{Ci}([1+c(M)]x)-\mathrm{Ci}(x)\right] -\frac{\sin(c(M)x)}{(1+c(M))x}\bigg\}\,.
 \end{eqnarray} 
We dropped the redshift dependence of the mass-concentration relation and $\tilde{u}_m$ and define $x = k R_{200}(M)/c(M)$, where $R_{200}$ is the cluster radius, and $\mathrm{Si}(x)$ and $\mathrm{Ci}(x)$ are the sine and cosine integrals.

%%%%%%%%%%%%%%%%%%%%%%%%%%%%%%%%%%%
%%%%%%%%%%%%%%%%%%%%%%%%%%%%%%%%%%%
\section{Simulated Likelihood Analysis - baseline scenario}
\label{sec:baseline}
%%%%%%%%%%%%%%%%%%%%%%%%%%%%%%%%%%%
%%%%%%%%%%%%%%%%%%%%%%%%%%%%%%%%%%%
We simulate an LSST like survey and summarize all parameters defining survey, cosmology, and systematics for our baseline scenario in Table \ref{tab:params}. 

\renewcommand{\arraystretch}{1.3}
\begin{table}
\caption{Fiducial parameters, flat priors (min, max), and Gaussian priors ($\mu$, $\sigma$)}
\begin{center}
\begin{tabular*}{0.45\textwidth}{@{\extracolsep{\fill}}| c c c |}
\hline
\hline
Parameter & Fid & Prior \\  
\hline 
\multicolumn{3}{|c|}{\tbf{Survey}} \\
$\Omega_{\mathrm{s}}$ & 18,000 deg$^2$ & fixed\\
$\sigma_\epsilon$ &0.26& fixed\\
\hline 
\multicolumn{3}{|c|}{\tbf{Cosmology}} \\
$\om$ & 0.3156 &  flat (0.1, 0.6)  \\ 
$\sig$ & 0.831 &  flat (0.6, 0.95)  \\ 
$\ns$ & 0.9645 & flat (0.85, 1.06)  \\
$\w$ &  -1.0 &   flat (0.0, 2.0)   \\
$\wa$ &  0.0 &  flat (-2.5, 2.5)   \\
$\omb$ &  0.0492 &  flat (0.04, 0.055)  \\
$h_0$  & 0.6727 &  flat (0.6, 0.76)   \\
\hline
\multicolumn{3}{|c|}{\tbf{Galaxy bias}} \\
$b_\mr{g}^1$ & 1.35  & flat (0.8, 2.0) \\
$b_\mr{g}^2$ & 1.5  &flat (0.8, 2.0) \\
$b_\mr{g}^3$ & 1.65 & flat (0.8, 2.0) \\
$b_\mr{g}^4$ & 1.8 & flat (0.8, 2.0) \\
\hline
\multicolumn{3}{|c|}{\tbf{Lens photo-z (red sequence)}} \\
$\Delta_\mr{z,lens}^i $ & 0.0 & Gauss (0.0, 0.0004) \\
$\sigma_\mr{z,lens} $ & 0.01 & Gauss (0.01, 0.0006) \\
\hline
\multicolumn{3}{|c|}{\tbf{Source photo-z}} \\
$\Delta_\mr{z,source}^i $ & 0.0 & Gauss (0.0, 0.002) \\
$\sigma_\mr{z,source}$ &0.05 & Gauss (0.05, 0.003) \\
\hline
\multicolumn{3}{|c|}{\tbf{Shear calibration}} \\
$m_i $ & 0.0 & Gauss (0.0, 0.004)\\
\hline
\multicolumn{3}{|c|}{\tbf{Cluster Mass Observable Relation}} \\
 $C_\lambda$ & 33.6 & 0.5 \\
 $\alpha_\lambda$ &1.08& 0.2 \\
 $\beta_\lambda$ &0.0& 0.5 \\
 $\sigma_{\mathrm{ln}M|\lambda}$ &0.25 & 0.2\\
 $c_\lambda^i$ &0.9 & 0.05\\
\hline
\end{tabular*}
\end{center}
\label{tab:params}
\end{table}
\renewcommand{\arraystretch}{1.0}

%%%%%%%%%%%%%%%%%%%%%%%%%%%%%%%%%%%
\subsection{Data vector}
\label{sec:dv}
%%%%%%%%%%%%%%%%%%%%%%%%%%%%%%%%%%%

\paragraph*{Source galaxies -- cosmic shear}
Adopting their `fiducial' galaxy selection cut, the true source redshift distribution $n_{\mathrm{source}}(z)$ is modeled as \citep{cjj13} 
\be 
\label{eq:redshiftben}
n_{\mathrm{source}}(z) \equiv \frac{d^2 N_{\mathrm{source}}}{dzd\Omega} = \bar{n}_{\mathrm{source}}\frac{ \Theta(z_{\mathrm{max}}-z)\; z^{1.24} \exp \left[ - \left(  \frac{z}{0.51} \right)^{1.01} \right]}{\int_0^{z_{\mathrm{max}}} dz\;z^{1.24} \exp \left[ - \left(  \frac{z}{0.51} \right)^{1.01} \right]} \,,
\ee 
imposing a high-$z$ cut $z_{\mathrm{max}}=3.5$, with $N_{\mathrm{source}}$ the total number of source galaxies, and $\bar{n}_{\mathrm{source}}$ the effective number density of source galaxies. After removal of masked and seriously blended objects, \citep{cjj13} find
\be
\label{eq:nsourcebar}
\bar{n}_{\mathrm{source}} = N_{\mathrm{source}}/\Omega_{\mathrm{s}} \approx 26\;\mathrm{galaxies/arcmin^2}\,.
\ee

This redshift distribution is then convolved with a photometric redshift uncertainty model, as described in Eqs.~(\ref{eq:photoz}, \ref{eq:redbin}) and split into ten tomographic bins, defined such than each photometric redshift bin contains the same number of galaxies. For the cosmic shear part of the data vector we compute 55 auto-and cross power spectra (see Sect. \ref{sec:sumstats}), which we divide into 21 logarithmically spaced Fourier mode bins ranging from $l_{\rm{min}} = 30$ to $l_{\rm{max}} = 5000$. 

\paragraph*{Lens galaxies -- clustering}
We assume a constant comoving redshift distribution for our lens sample \citep[similar to][]{rra15} with a projected number density $n_{\mathrm{lens}} =  0.25\; \mathrm{galaxies/arcmin^2}$ and divide this sample into 4 narrow redshift bins (0.2-0.4,0.4-0.6,0.6-0.8,0.8-1.0). The data vector is divided into 25 $l$-bins ranging from 30-15000, however we exclude high $l-$bins, if scales below $R_\mr{\min}=k_\mr{max}/2 \pi=10 \mr{ Mpc/h}$ contribute to the projected integral (see Eq. \ref{eq:projected}). In Sect. \ref{sec:Rvary} we vary the choice of $R_\mr{\min}$ and choose larger values, i.e. 20 Mpc/h and 50 Mpc/h, as well as a small $R_\mr{\min}=0.1 \mr{ Mpc/h}$. The latter requires extending our linear bias formalism to a Halo Occupation Distribution (HOD) model.  

\paragraph*{Photometric BAOs} We include 4 measurements of photometric BAOs at z=\{1.25, 1.5, 1.75, 2.0\} which enter the analysis as an independent probe. This is an approximation but likely justified since the strongest correlation of BAOs would show in the galaxy clustering signal, which is limited to lower redshifts. We assume that these photometric BAO measurements have an error of $\sigma_\mr{BAO}=0.03$.

\paragraph*{Lens $\times$ source galaxies -- galaxy-galaxy lensing}
The galaxy-galaxy lensing part of the data vector assumes the lens galaxy sample as foreground and the source galaxy sample as background galaxies; we only consider non-overlapping source and lens in redshift bins. We again impose a cut-off at $R_\mr{\min}=10 \mr{ Mpc/h}$ for the baseline model, which is varied in Sect. \ref{sec:Rvary} in accordance with the clustering part of the data vector. 

\paragraph*{Galaxy cluster number counts} We consider four cluster redshift bins (0.2-0.4,0.4-0.6,0.6-0.8, and 0.8-1.0) and seven cluster richness bins above $\lambda_{\mathrm{min}} = 10$ in each redshift bin.

\paragraph*{Galaxy clusters $\times$ source galaxies -- cluster weak lensing} In order to calibrate the cluster mass--richness relation (Eq.~\ref{eq:Mlambda}), we consider the stacked weak lensing signal from all combinations of cluster redshift and richness bins $\delta_{\lambda_{\alpha}}^i$ with source galaxies $\kappa^j$, with the restriction that source galaxies are located at higher redshift than the galaxy clusters. Specifically, we use the cluster lensing power spectrum in the angular range $3000 < l <15000$, which corresponds mostly to the 1-halo cluster lensing signal. We note that large-scale cluster lensing using clusters in different richness bins would be a prime candidate for a multi-tracer cosmology \cite{sel09} approach, but we postpone a discussion of constraints from large-scale cluster lensing (and cluster clustering) to future work, as this requires a detailed examination of additional systematic uncertainties such as assembly bias and stochasticity.

\begin{figure*}
  \begin{minipage}[c]{0.85\textwidth}
\includegraphics[trim={1.8cm 0.4cm 2cm 1.2cm},clip,width=\textwidth]{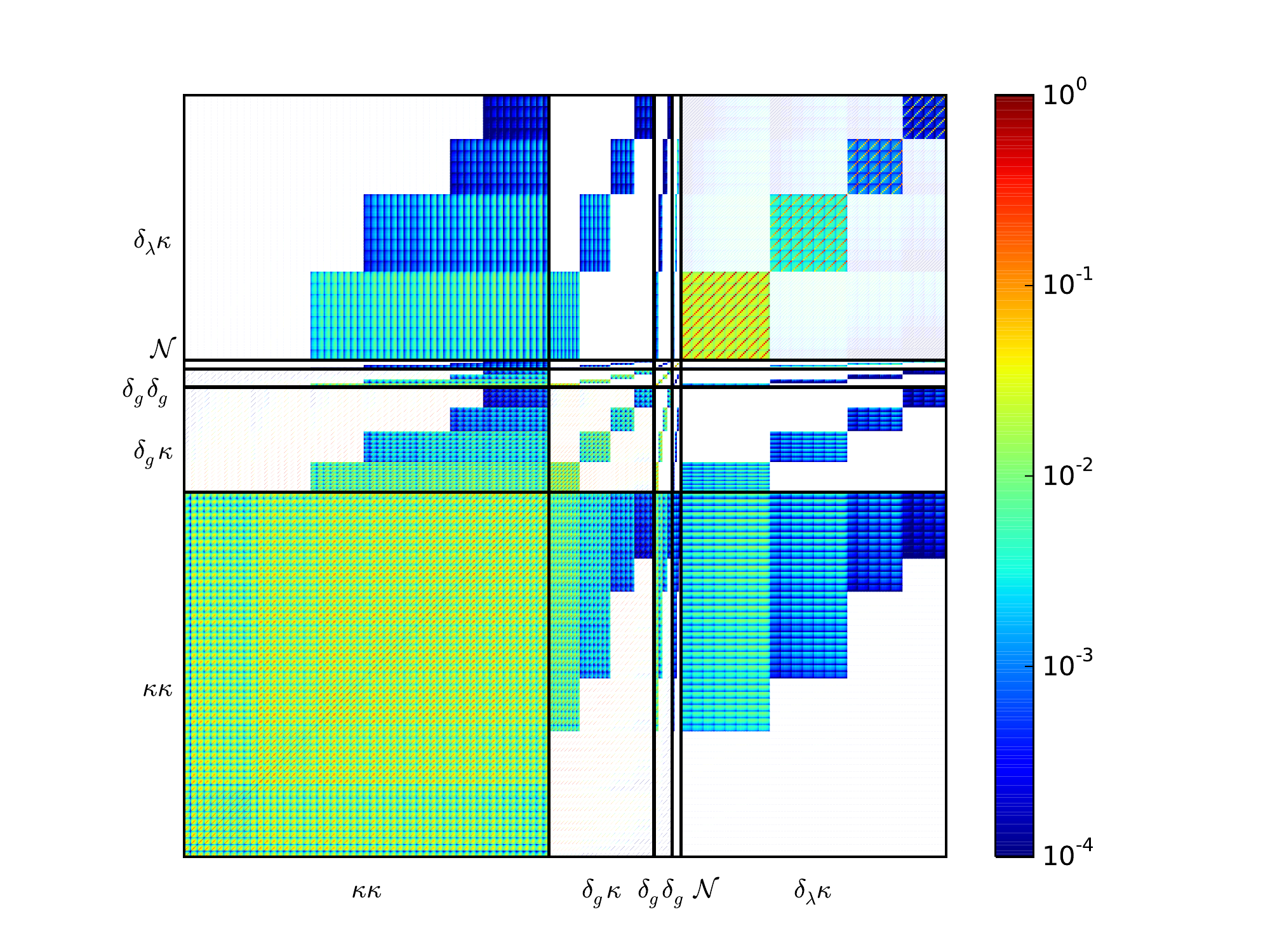}
  \end{minipage}\hfill
  \begin{minipage}{0.15\textwidth}
   \caption{Non-Gaussian, multi-probe correlation matrix for a joint data vector of cosmic shear, galaxy-galaxy lensing, galaxy clustering, cluster number counts, and cluster weak lensing. Details on the calculation of this $\sim$ 5.8 million entry matrix (7.4 million when going to the smallest scales considered in this paper) can be found in Appendix \ref{sec:app1}. We recommend a zoom factor of $\sim$10 to gain more insight into the matrix structure, e.g. to actually identify individual elements.} \label{fi:covstruct}
  \end{minipage}
\end{figure*}

\subsection{Systematics} 
\label{sec:sys}
We parameterize uncertainties arising from systematics through nuisance parameters, which are summarized with their fiducial values and priors in Table \ref{tab:params}. Our default likelihood analysis includes the following systematics:

\paragraph*{Photometric redshift uncertainties} As described in detail in \citet{mhh06}, the true redshift distribution of galaxy population $x$ (here, $x\in\{\mathrm{lens},\; \mathrm{source}\}$) in photometric redshift bin $i$ with $z^i_{\mathrm{ph,min},x} <z_{\mathrm{ph}}< z^i_{\mathrm{ph,min},x}$ can be written as
\be
\label{eq:photoz}
n^i_x(z) = \int_{z^i_{\mathrm{ph,min},x}}^{z^i_{\mathrm{ph,max},x}} dz_{\mathrm{ph}} p^i\left(z_{\mathrm{ph}}|z,x\right)\,,
\ee
where $p\left(z_{\mathrm{ph}}|z,x\right)$ is the probability distribution of $z_{\mathrm{ph}}$ at given true redshift $z$ for galaxies from population $x$. 
Furthermore, the number density of galaxies in this redshift bin is given by
\be
\label{eq:nbar}
\bar{n}_x^i = \int dz\; n_x^i(z).
\ee

In this analysis we only consider Gaussian photometric redshift uncertainties, which are characterized by scatter $\sigma_z(z)$ and bias $\Delta_z(z)$. While these may in general be arbitrary functions, we further assume that the scatter can be described by the simple redshift scaling $\sigma_{z,x}(1+z)$ and allow one (constant) bias parameter $\Delta^i_{z,x}$ per redshift bin:
\be
\label{eq:redbin}
p^i\left(z_{\mathrm{ph}}|z,x\right) = \frac{1}{\sqrt{2\pi}\sigma_{z,x}(1+z)}
\exp\left[-\frac{\left(z-z_{\mathrm{ph}} - \Delta^i_{z,x}\right)^2}{2\left(\sigma_{z,x}(1+z)\right)^2}\right]\,.
\ee
For our four lens galaxy redshift bins, this model results in five parameters (four photo-z biases, and one photo-z scatter parameter); and in 11 additional parameters for the 10 source galaxy redshift bins. The fiducial values of $\sigma_z(z)$ and bias $\Delta_z(z)$ including their priors for the source sample correspond to the LSST requirements. The values for the lens sample are based on experience from the DES survey, where red sequence galaxies show a similar improvement over DES source galaxies. We note that this level of photo-z accuracy requires improvements in the corresponding measurement techniques over current standards (for the assumed lens sample accuracy, peculiar velocities are starting to be an effect).   

\paragraph*{Linear galaxy bias} is described by one nuisance parameter per lens galaxy redshift bin, which is marginalized over using conservative flat priors.
\paragraph*{Multiplicative shear calibration} is modeled using one parameter $m^i$ per redshift bin, which affects cosmic shear and galaxy-galaxy lensing power spectra via
\bea
\nonumber C_{\kappa \kappa}^{ij}(l) \quad &\longrightarrow& \quad (1+m^i) \, (1+m^j) \, C_{\kappa \kappa}^{ij}(l), \\
C_{\delta_{\mathrm{g}}\kappa}^{ij}(l) \quad &\longrightarrow& \quad (1+m^j) \, C_{\delta_{\mathrm{g}}\kappa}^{ij}(l),
\eea
where the cluster lensing power spectra are affected analogously to the galaxy-galaxy lensing spectra. We marginalize over each $m^i$ independently with Gaussian priors (10 parameters).

\paragraph*{Cluster mass-observable relation} The fiducial values for the mass-richness relation parameters $(C_\lambda,\alpha_\lambda)$ are adopted from Eq.~(B4)  in \citet{rkr12}, transformed to $h_{100}$ units; we furthermore use 
$\beta_\lambda = 0$ and $\sigma_{\mathrm{ln}M|\lambda} = 0.25$ as baseline value. We marginalize over these parameters using flat priors, which are agnostic to previous measurements of the mass-richness relation.

In addition, we use a simplified model for the cluster selection incompleteness with one parameter $c_\lambda^i$ per redshift bin,
\be 
\mathcal{N}^i(\lambda_\alpha) \rightarrow c_\lambda^i \mathcal{N}^i(\lambda_\alpha)\,,
\ee
for which we assume fiducial values $c_\lambda^i = 0.9$ and Gaussian priors, in conjunction with a flat prior $c_\lambda^i \le 1$.

\paragraph*{Intrinsic galaxy alignment} This is not part of our baseline analysis but will be considered in Sect. \ref{sec:IA}. In short we follow \cite{keb16} but extend the IA formalism to galaxy-galaxy lensing. 

\paragraph*{Other systematics} There are several important sources of systematic uncertainties to be considered in future extensions of this work. For example, baryonic effects and other modeling uncertainties on small scales of projected power spectra must be accounted for. In \cite{ekd15} we have developed a mitigation technique for baryonic effects in cosmic shear that removes corresponding LSST biases effectively even out to $l=5000$ \citep[also see][]{zsd13,mph15}. This idea should be extended to all probes considered in this paper. We also postpone implementing galaxy cluster mis-centering, assembly bias and stochasticity.

%%%%%%%%%%%%%%%%%%%%%%%%%%%%%%%%%%%
\subsection{Likelihood formalism} 
\label{sec:like}
%%%%%%%%%%%%%%%%%%%%%%%%%%%%%%%%%%%
\begin{figure*}
  \includegraphics[width=17cm]{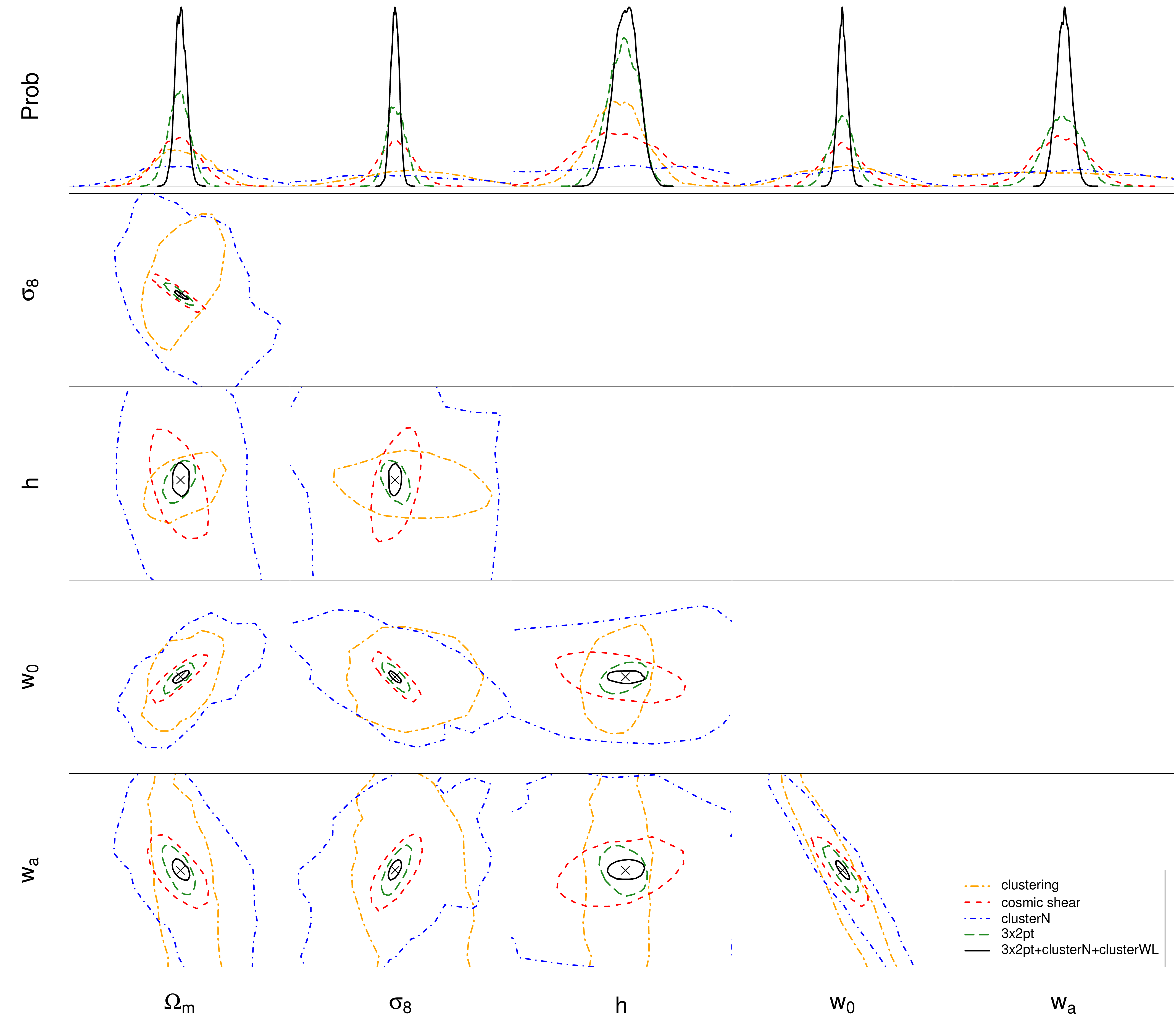}
\caption{Individual vs. multi-probe cosmological constraints. We show projected cosmological constraints for clustering \ti{(orange/dot-long dashed)}, cosmic shear \ti{(red/dashed)}, cluster number counts \ti{(blue/dot-dashed)} individually. The 3x2pt multi-probe contours \ti{(green/long-dashed)} include information from clustering, cosmic shear, and galaxy-galaxy lensing; the \ti{black/solid} contours add information from cluster number counts and cluster weak lensing to the 3x2pt data vector, altogether 2413 data points.}
 \label{fi:moneyplot}
\end{figure*}

The multi-probe data vector, denoted as $\D$, is computed at the fiducial parameters in cosmology and systematics see Table \ref{tab:params}. The same parameters enter in the computation of the non-Gaussian covariance matrix $\matC$. We show the corresponding correlation matrix in Fig. \ref{fi:covstruct} and detail the calculation of the individual terms in Appendix \ref{sec:app1}. We note that the computation and verification of this matrix was the most time-consuming aspect of this paper. Since this covariance matrix is calculated analytically and not estimated from either simulations or data, it does not inherently limit the number of data points that can enter our analysis \citep[see e.g.,][for details on these constraints]{hss07,tjk13,dos13}. 

We sample the joint parameter space of cosmological $\pco$ and nuisance parameters $\pnu$ and parameterize the joint likelihood as a multivariate Gaussian 
\be
\label{eq:like}
\like (\D| \pco, \pnu) = N \, \times \, \exp \biggl( -\frac{1}{2} \underbrace{\left[ (\D -\M)^t \, \matC^{-1} \, (\D-\M) \right]}_{\chi^2(\pco, \pnu)}  \biggr) \,.
\ee
The model vector $\M$ is a function of cosmology and nuisance parameters, i.e. $\M=\M(\pco, \pnu)$ and the normalization constant $N=(2 \pi)^{-\frac{n}{2}} |C|^{-\frac{1}{2}}$ can be ignored under the assumption that the covariance is constant in parameter space. The assumption of a constant, known covariance matrix $\matC$ is an approximation to the correct approach of a cosmology dependent or estimated covariance \citep[see][for further details]{esh09, seh15}. We examine the impact of the covariances' input cosmology on likelihood contours in Sect. \ref{sec:covvary}.
    
Given the likelihood function we can compute the posterior probability in parameter space from Bayes' theorem
\be
\label{eq:bayes}
\prob(\pco, \pnu|\D) \propto \probr (\pco, \pnu) \,\like (\D| \pco, \pnu),
\ee
where $\probr (\pco, \pnu)$ denotes the prior probability (non-informative priors for the case of this paper).

%%%%%%%%%%%%%%%%%%%%%%%%%%%%%%%%%%%
\subsection{Results - baseline scenario}
\label{sec:resbase}
%%%%%%%%%%%%%%%%%%%%%%%%%%%%%%%%%%%
 
Results of our baseline LSST likelihood analysis simulation are shown in Fig. \ref{fi:moneyplot}. All contours include systematic effects that are associated with the corresponding probe(s). Correspondingly, the dimensionality of the likelihood analyses differs substantially; it ranges from 15 for the cluster number count analysis to 45 for the joint analysis of all 5 probes considered in the data vector.  
 
We find that the galaxy clustering analysis with the imposed cut-off scale of $R_\mr{min}=10.0 \mr{Mpc/h}$ is strongly affected by systematics, most likely our unconstrained galaxy bias. Cosmic shear in itself has relatively tight constraints, however we see a substantial increase when combining the two aforementioned probes with galaxy-galaxy lensing (denoted as 3x2pt). 

Whereas cluster number counts alone gives the weakest constraints overall, it is extremely promising when combining it with the 3x2pt scenario and adding cluster weak lensing to calibrate cluster masses. The information gain from 3x2pt to the scenario where all probes are included is remarkable. One reason is the fact that clusters contribute small scale clustering information from the 1H-term, which is not present in the clustering or galaxy-galaxy lensing data (also see Sect. \ref{sec:Rvary}). Another reason to caution against overestimating the effect of clusters is the fact that we have not yet considered galaxy cluster mis-centering, assembly bias and stochasticity as additional uncertainties.

Combining multiple probes has a highly non-linear effect on cosmological constraining power. It should be an important aspect of future work to explore optimal multi-probe data vectors for the various science cases (beyond cosmic acceleration).

%%%%%%%%%%%%%%%%%%%%%%%%%%%%%%%%%%% 
%%%%%%%%%%%%%%%%%%%%%%%%%%%%%%%%%%%
\section{Scenarios beyond the baseline analysis}
\label{sec:applications}
%%%%%%%%%%%%%%%%%%%%%%%%%%%%%%%%%%%
%%%%%%%%%%%%%%%%%%%%%%%%%%%%%%%%%%%
In this section we illustrate some of the \textsc{CosmoLike} capabilities to forecast and optimize the LSST survey. Starting out from the baseline model we vary the galaxy lens and source samples as well as associated systematics. We also examine constraints when including highly non-linear scales in the lens sample, which requires us to replace the linear galaxy bias computation with \textsc{CosmoLike}'s HOD module. We also vary the input cosmology of the computed covariance matrix as a first step to quantify the impact of this choice on cosmological constraints. Lastly, we consider the impact of galaxy intrinsic alignment for the multi-probe case and in the presence of multiple systematics. 
 
\subsection{Varying galaxy samples: systematics vs. statistics}
\label{sec:galvary}

\begin{figure}
  \includegraphics[width=8.5cm]{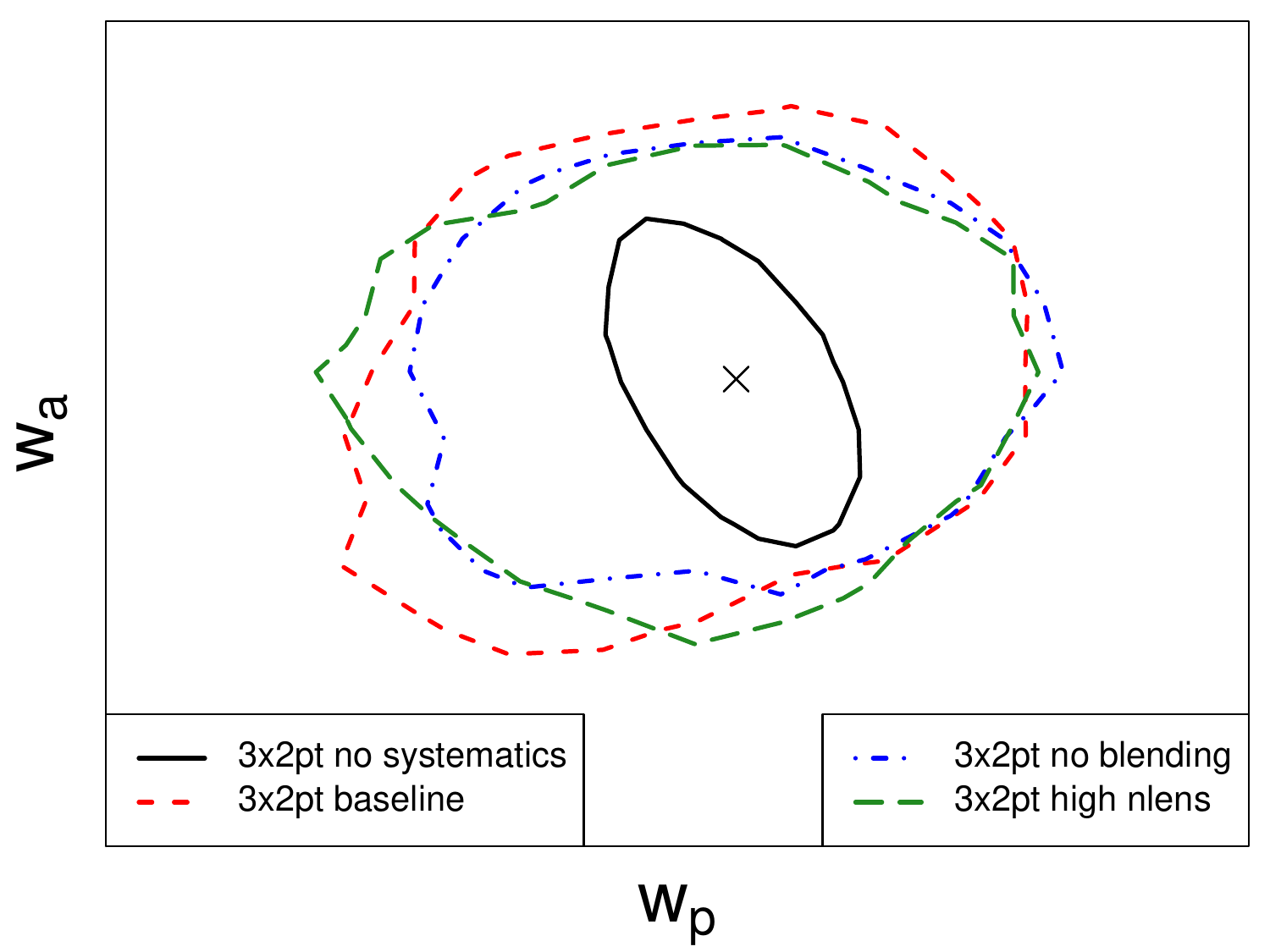}
\caption{Impact of galaxy samples and associate systematics on cosmological information. We show the systematics free 3x2pt function case \ti{(black, solid)} in comparison to our baseline model \ti{(red/dashed)}. The \ti{(blue, dot-dashed)} contours show the information gain when including all blended objects in the analysis, i.e. increasing $\bar{n}_\mr{source}$ from 26 to 37 galaxies/arcmin$^2$; \ti{green/long-dashed} constraints are obtained when including a lens galaxy sample that is by a factor of 20 larger than our baseline (red sequence) sample, but has worse photo-z accuracy.}
 \label{fi:galvary}
\end{figure}

Statistical power of photometric surveys comes from covered area, to reduce cosmic variance, and from the number density of galaxies, to reduce noise contributions when estimating summary statistics. Maximizing the number density of galaxies requires the inclusion of faint, small, and poorly understood  galaxies, which give rise to additional systematics. The trade-off between statistical power and systematics needs to be simulated carefully to select optimal galaxy samples and to focus future research on the most limiting factors of an analysis. 
  
Figure \ref{fi:galvary} illustrates the difference in cosmological information when comparing a systematics-free 3x2pt analysis \ti{(black/solid)} to our baseline scenario \ti{(red/dashed)} that includes uncertainties from photo-z's, shear calibration, and galaxy bias (see Table \ref{tab:params}). 

The main contributors in reducing source galaxies for LSST are masking and atmospheric blending \citep{cjj13,dst16}. For example, \citep{cjj13} find that these effects shrink the number density of source galaxies from 37 to 26 galaxies/arcmin$^2$. The \ti{(blue/dot-dashed)} contours show results of a simulated analysis assuming 37 galaxies/arcmin$^2$. Since we do not assume an increase in photo-z and shear calibration uncertainties, these contours correspond to an upper limit in information gain when solving the problem of blending for LSST.

The \ti{(green/dashed)} contours illustrate results when considering a lens galaxy sample that has a factor of 20 higher number density of galaxies compared to our baseline scenario, but degraded photo-z accuracy (compare Tables \ref{tab:params} and \ref{tab:params2}).

We find very limited gain in information when increasing the number density of either source or lens galaxies, which we explain as follows: First, our error budget is systematics dominated (indicated by \ti{black/solid} vs \ti{red/dashed contours}). Second, the Non-Gaussian cosmic variance terms in our covariance matrix likely dominate the noise contributions; increasing the number density of galaxies and hence decreasing the noise has no effect. An increase in survey area (e.g., towards the equator, which would also allow for increased overlap with the Dark Energy Spectroscopic Instrument survey) would be a more promising approach.

\renewcommand{\arraystretch}{1.3}
\begin{table}
\caption{Parameters, flat priors (min, max), and Gaussian priors ($\mu$, $\sigma$) for non-baseline scenarios considered in Sect. \ref{sec:applications}}
\begin{center}
\begin{tabular*}{0.45\textwidth}{@{\extracolsep{\fill}}| c c c |}
\hline
\hline
Parameter & Fid & Prior \\  
\hline
\multicolumn{3}{|c|}{\tbf{High density lens sample considered in Fig. \ref{fi:galvary}}} \\
 $\Delta_\mr{z,lens}^i $ & 0.0 & Gauss (0.0, 0.001) \\
$\sigma_\mr{z,lens} $ & 0.04 & Gauss (0.04, 0.002) \\
 \hline
\multicolumn{3}{|c|}{\tbf{HOD implementation in Fig. \ref{fi:HOD}}} \\
$M_{\mr{min}}$ &12.1 & flat (10,15)\\
$M_1^{\prime}$ &13.65 & flat (10,15)\\
$M_0$ &12.2 & flat (10,15)\\
$\sigma_{\mr{ln}M}$ & 0.4  & flat (0.1,1.0) \\
$\alpha_{\mr{sat}}$ & 1.0  &  flat (0.5,1.5) \\
$f_{\mr{c}}$ & 0.25  &  flat (0.1,1.0)  \\
 \hline
\multicolumn{3}{|c|}{\tbf{Covariance cosmology changes in Fig. \ref{fi:covvary}, model1}} \\
 $\om $ & 0.284 & no prior - fixed value \\
 $\sig $ & 0.748 & no prior - fixed value \\
 \hline
\multicolumn{3}{|c|}{\tbf{Covariance cosmology changes in Fig. \ref{fi:covvary}, model2}} \\
$\w $ & -1.3 & no prior - fixed value \\
 $\wa $ & -0.5 & no prior - fixed value \\
 \hline

\end{tabular*}
\end{center}
\label{tab:params2}
\end{table}
\renewcommand{\arraystretch}{1.0}

%%%%%%%%%%%%%%%%%%%%%%%%%%%%%%%%%%%
\subsection{Varying $R_\mr{min}$: linear galaxy bias vs. HOD model}
\label{sec:Rvary}
%%%%%%%%%%%%%%%%%%%%%%%%%%%%%%%%%%%
In this subsection we address the change of information content as a function of scale to which galaxy biasing can be modeled accurately. Our baseline scenario includes cosmic shear up to $l_\mr{max}=5000$, however it imposes an $R_\mr{min}=10\mr{ Mpc/h}$ cut-off for clustering and galaxy-galaxy lensing. 
Perturbative models for galaxy biasing in the quasi-linear regime is an active area of research \citep[e.g.][]{McDR09,Senatore15,AFSV15}, 
and the model for galaxy clustering and galaxy-galaxy lensing in Eq.~(\ref{eq:Pg}) needs to be updated for analyses of galaxy clustering measurements from future surveys. However, in the context of this forecast study, we are primarily interested in cosmological information content as a function of scale. Forecasts based on the effective linear biasing model should be interpreted as the potential constraining power assuming that sufficiently accurate bias models will be developed by the time of the data analysis. 

First, we characterize the loss in cosmological information from more conservative $R_\mr{min} = 20\mr{ Mpc/h}$ and $R_\mr{min} = 50\mr{ Mpc/h}$. Second, we consider a very optimistic scenario, in which we assume that galaxy biases down to scales of $R_\mr{min} = 0.1\mr{ Mpc/h}$ and over the redshift range $0.2 < 0.8$ can be described by a simple non-linear model.

\begin{figure*}
 \includegraphics[width=8.5cm]{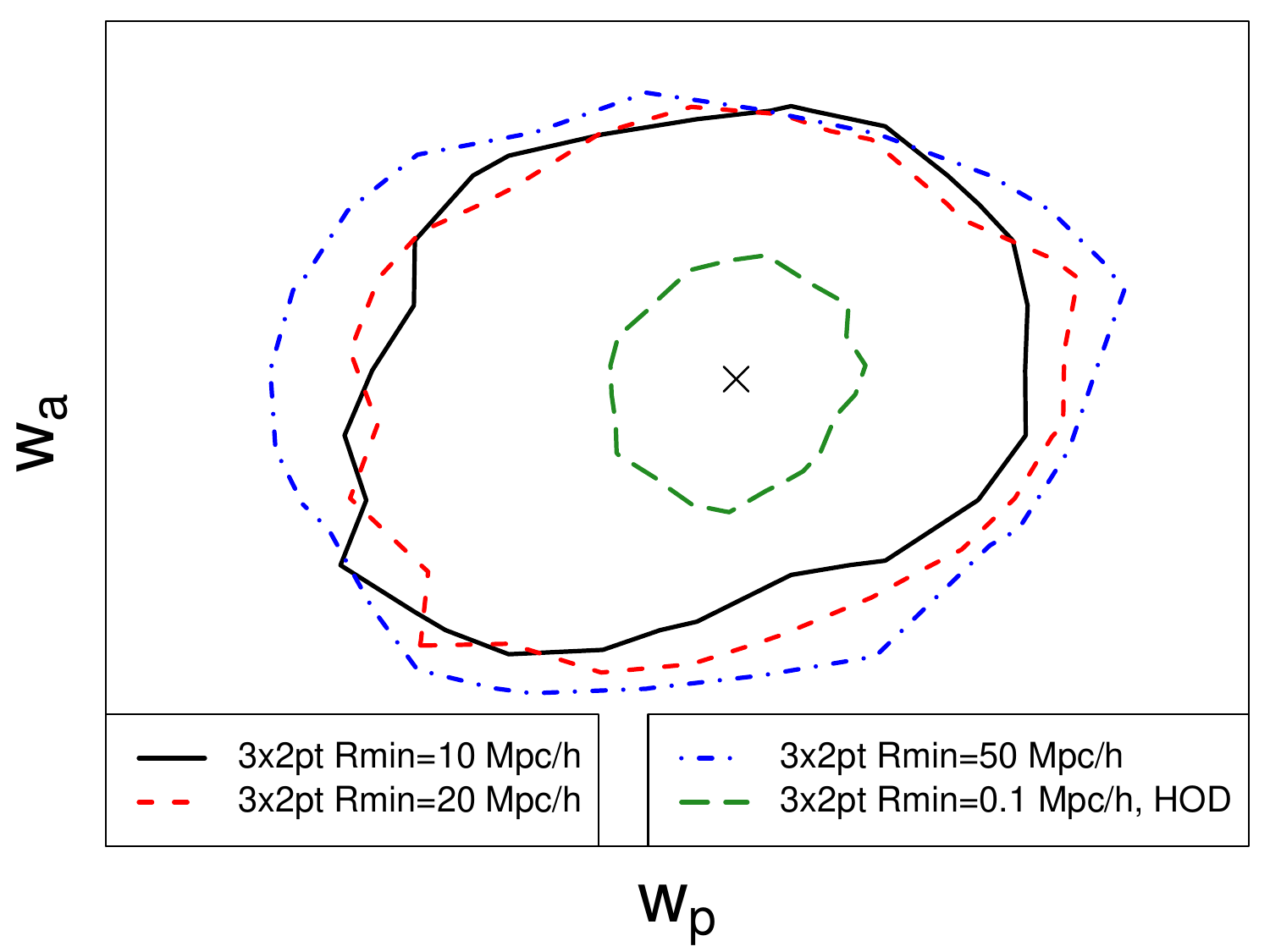}
 \includegraphics[width=8.5cm]{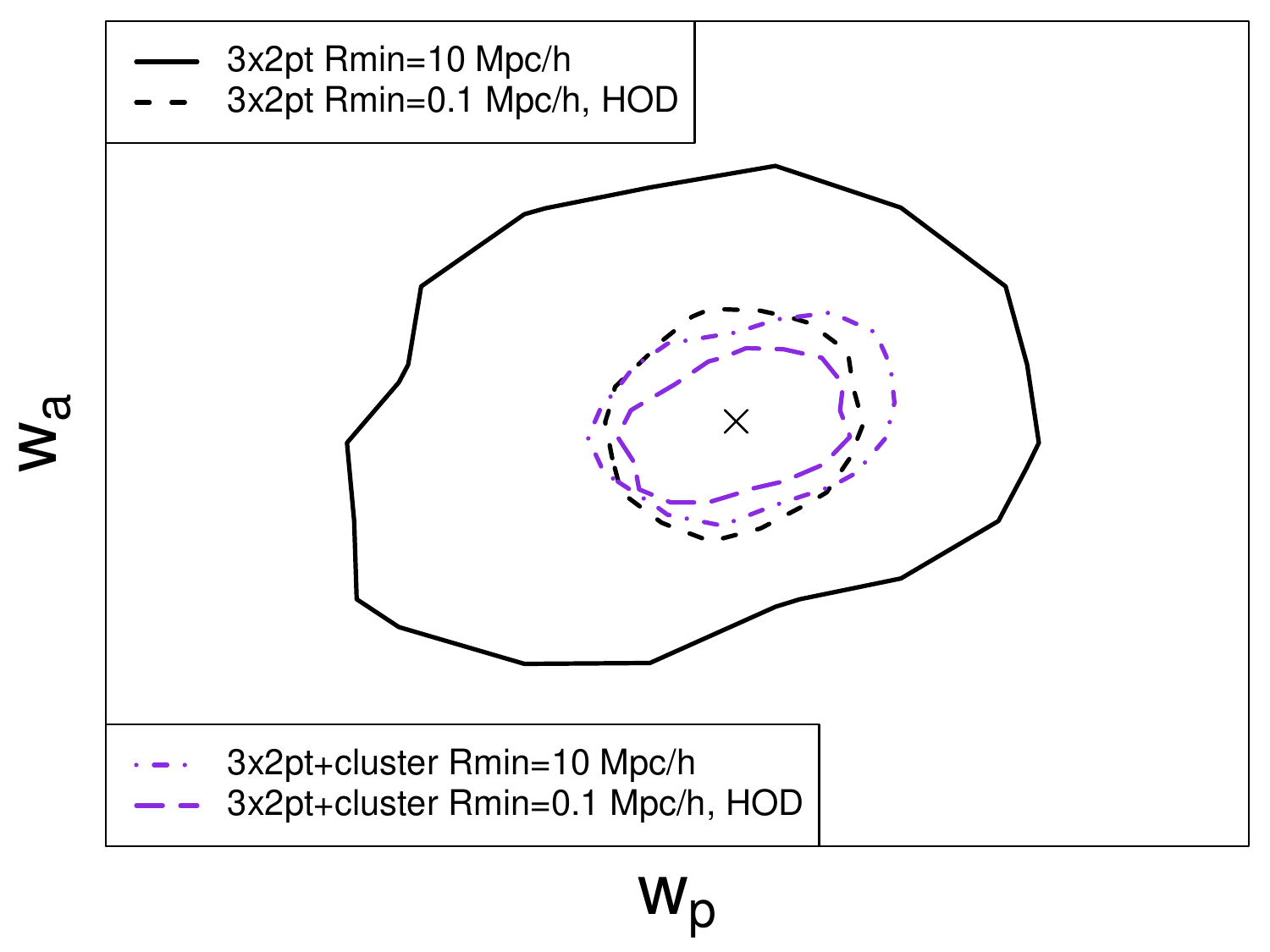}
\caption{\tbf{Left}: Varying the minimum scale included in galaxy clustering and galaxy galaxy lensing measurements. We show the baseline 3x2pt functions, which assumes $R_\mr{min}=10\mr{ Mpc/h}$ \ti{(black/solid)}, and corresponding constraints when using $R_\mr{min}=20 \mr{Mpc/h}$ \ti{(red/dashed)}, $R_\mr{min}=50 \mr{Mpc/h}$ \ti{(blue/dot-dashed)}, $R_\mr{min}=0.1 \mr{Mpc/h}$ \ti{(green/long-dashed)} instead. For the latter we switch from linear galaxy bias modeling to our HOD implementation. \tbf{Right:}  Information gain when using HOD instead of linear galaxy bias for 3x2pt \ti{(black solid vs dashed contours)} in comparison to corresponding information gain when including cluster number counts and cluster weak lensing in the data vector \ti{(violett/dot-dashed vs long-dashed)}.}
 \label{fi:HOD}
\end{figure*}
For the latter, we replace Eq.~(\ref{eq:Pg}) by a Halo Occupation Distribution (HOD) model 
\citep[e.g.][]{BW02,zbw05,vdB12}, which describe the relation between galaxies and mass in terms of the probability that a halo of given mass contains $N_{\mathrm{g}}$ galaxies. Following \citet{zbw05}, we split the HOD into central and satellite terms, which we model as \citep{zwb11}
\bea
\nonumber \ensav{N_\mathrm{c}(M)} =& \frac{1}{2}\left[1+\mr{erf}\left(\frac{\log M-\log M_{\mr{min}}}{\sigma_{\mr{ln}M}}\right)\right] \,,\\
\ensav{N_\mathrm{s}(M)} =& \Theta(M-M_0) \left(\frac{M-M_0}{M_1^{\prime}}\right)^{\alpha_{\mathrm{sat}}}\,.
\eea
The central occupation is a softened step function with transition mass $M_{\mr{min}}$, which the characteristic mass or a halo to host a central galaxy, and softening $\sigma_{\mr{ln}M}$. $M_1^\prime$ is the characteristic mass scale for a halo to have a satellite galaxy; the satellite distribution is a power law with slope $\alpha_{\mathrm{sat}}$ in high mass halos, and it is cut off at a low mass scale $M_0$. For luminosity threshold samples, the satellite occupation is typically modulated by the central galaxy occupation, as a halo has to contain a central galaxy to have satellite galaxies. For a color selected sample however, only a fraction $f_\mathrm{c}$ meets the sample selection criteria, and we write the total galaxy occupation as
\be
\ensav{N_\mathrm{g}(M)} = \ensav{N_{\mathrm{c}}(M)}\Big[f_{\mathrm{c}} + \ensav{N_{\mathrm{s}}(M)}\Big]\,.
 \label{eq:NM}
\ee
Based on this HOD, we calculate the galaxy-galaxy lensing and clustering power spectra as 
\bea
\label{eq:PHOD}
\nonumber P_{\mathrm{gm}} (k,z)& = & b_{\mathrm{HOD}}(z) P_{\mathrm{lin}}(k,z)\\
\nonumber& +& \frac{\int dM \frac{dn}{dM}M u_{\mathrm{m}}(k,M) \ensav{N_{\mathrm{c}}(M)}\ensav{N_{\mathrm{s}}(M)}u_{\mathrm{s}}(k,M)}{\bar{\rho}\int dM\frac{dn}{dM}\ensav{N_\mathrm{g}(M)}}\\
\nonumber P_{\mathrm{gg}} (k,z) &=& \left(b_{\mathrm{HOD}}(z)\right)^2 P_{\mathrm{lin}}(k,z)\\
&+&\frac{\int dM \frac{dn}{dM}\ensav{\Big\{
N_{\mathrm{c}}(M)\left[f_{\mathrm{c}} + N_{\mathrm{s}}(M)\tilde{u}_{\mathrm{s}}(k,M)\right)]\Big\}^2}}
{\left(\int dM \frac{dn}{dM}\ensav{N_\mathrm{g}(M)}\right)^2}\,
\eea
with $\tilde{u}_{\mathrm{s}}(k,M)$ the Fourier transform of the satellite galaxy density profile, which we assume to follow the matter density profile, and where for notational convenience we define $\ensav{\left[N_{\mathrm{c}}(M)\right]^2} \equiv 0$.

The left panel of Fig.  \ref{fi:HOD} shows a tolerable loss in information when going from the baseline 3x2pt scenario ($R_\mr{min}=10 \mr{Mpc/h}$, \ti{black/solid}) to even larger cut-offs such as $R_\mr{min}=20 \mr{Mpc/h}$, \ti{(red/dashed)} and $R_\mr{min}=50 \mr{Mpc/h}$, \ti{(blue/dot-dashed)}. This  is in sharp contrast to the substantial information gain when employing \textsc{CosmoLike}'s HOD module to include smaller scales ($R_\mr{min}=0.1 \mr{Mpc/h}$, \ti{green/long-dashed}) in the analysis. The same information gain however is less significant when adding cluster number counts and cluster weak lensing to the 3x2pt data vector (right panel). A likely explanation is the fact that clusters themselves are highly sensitive to small, nonlinear scales and corresponding information from galaxy clustering and galaxy-galaxy lensing is somewhat redundant.  

%%%%%%%%%%%%%%%%%%%%%%%%%%%%%%%%%%%
\subsection{Varying cosmology in covariances}
\label{sec:covvary}
%%%%%%%%%%%%%%%%%%%%%%%%%%%%%%%%%%%
\begin{figure}
  \includegraphics[width=8.5cm]{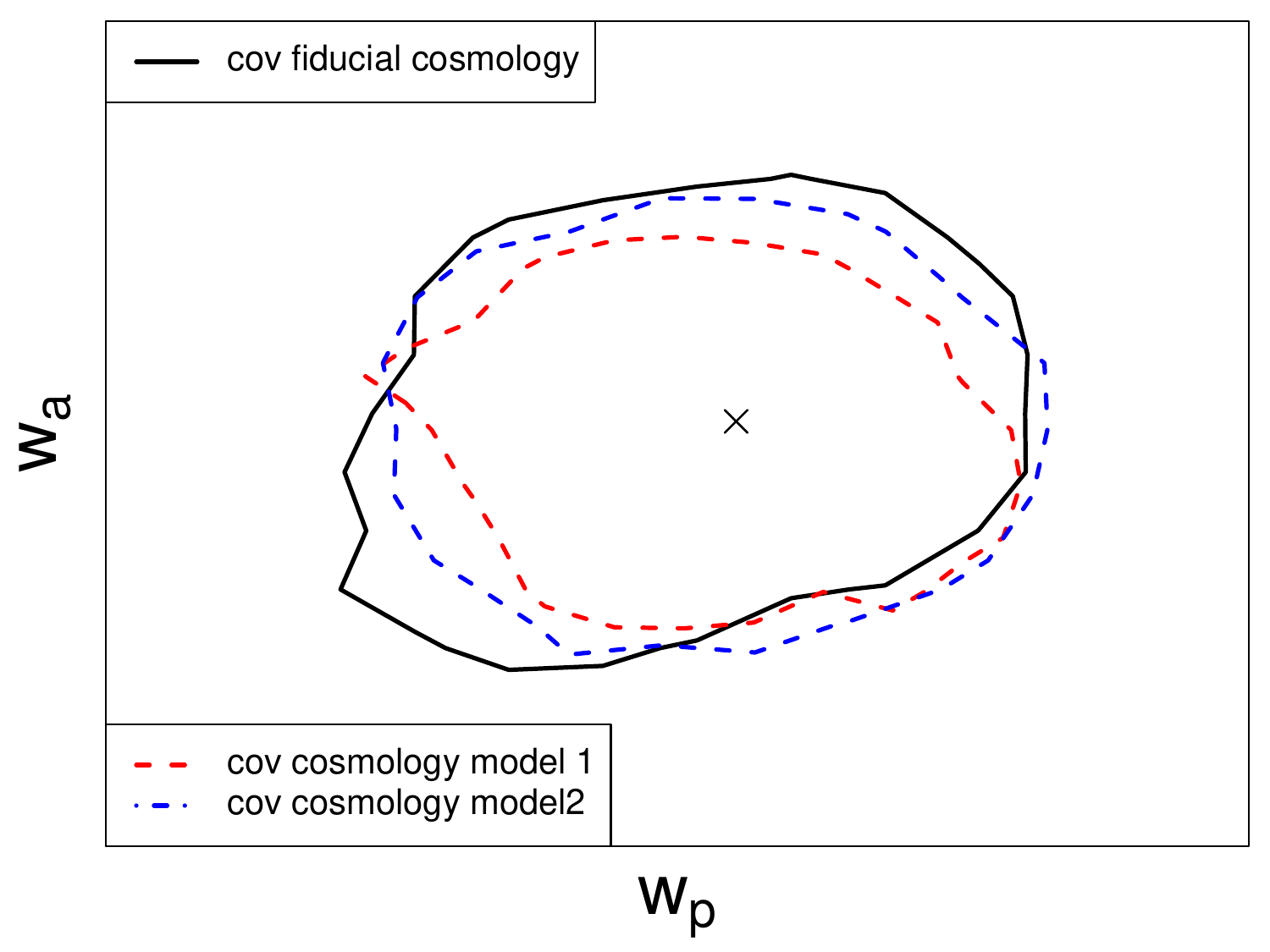}
\caption{Change in cosmological constraints when varying the underlying cosmological model in the covariance matrix. We show three scenarios: 1) the fiducial cosmology (black/solid), 2) fiducial cosmology but a 10\% lower value in $\sig$ and $\om$ \ti{(red/dashed)}, and 3) fiducial cosmology but changes in the dark energy parameters, i.e. $\w=-1.3$ and $\wa=-0.5$ \ti{(blue/dot-dashed)}.}
 \label{fi:covvary}
\end{figure}

Covariance matrices pose a major obstacle in multi-probe cosmological analyses. If they are obtained through (re)sampling methods using the data itself they are an estimated quantity (similar to the estimated data vector), which changes the functional form of the likelihood from a multivariate Gaussian to a modified multivariate $t$-distribution \citep{seh15}. If the covariance is computed analytically it must be considered a `known quantity' that follows deterministically from the cosmological (and nuisance parameters) that are under evaluation. Consequently, the covariance must vary in accordance with the sampler walking through the parameter space \citep{esh09}. If the covariance is estimated from a set of simulations it is both an estimated quantity and it assumes an underlying cosmology, namely that of the simulation. Technically any analysis using such a covariance matrix requires a combination of repeatedly computing the covariance as a function of the parameters considered and adopting a $t$-distribution as the functional form of the likelihood. 
   
In practice, the fact that the covariance matrix is an estimated quantity is generally ignored when inferring cosmological parameters; similarly most analyses ignore the covariance's cosmology dependence \citep[a notable exception is][who conduct a non-tomographic cosmic shear analysis]{jts13}. 

In Fig. \ref{fi:covvary} we simulate 3 likelihood analyses for a 3x2pt data vector 	that assume analytically calculated covariance matrices with different underlying cosmologies. We find that when reducing the fiducial values for $\om$ and $\sig$ by$\sim 10\%$ \ti{(red/dashed)}, the contours shrink moderately. This is expected since lower amplitude in $\om$ and $\sig$ reduce cosmic variance terms in the covariance. We find only minor changes in the contours when decreasing our fiducial dark energy parameters to $\w=-1.3$ and $\wa=-0.5$ \ti{(blue/dot-dashed)}.  

This initial study needs a more thorough follow up analysis \citep[e.g., by implementing the method suggested in][for multi-probe covariances]{mos13} for several reasons: first, we have only checked for the impact of the covariances' cosmological model in the context of the standard likelihood technique, i.e. fixing the covariance cosmology throughout the MCMC walk. Fully accounting for the cosmology dependence implies that the likelihood function's exponential term and its normalization change continuously as a function of parameter space, which can be a stronger effect than the one examined in Fig. \ref{fi:covvary}. Second, the change of the covariance matrix with respect to nuisance parameters has not been examined to date. Third, contours in Fig. \ref{fi:covvary} marginalize over 30 nuisance parameters, which washes out differences that can become more severe if systematics control improves.

%%%%%%%%%%%%%%%%%%%%%%%%%%%%%%%%%%%
\subsection{Intrinsic alignment with multiple probes and systematics}
\label{sec:IA}
%%%%%%%%%%%%%%%%%%%%%%%%%%%%%%%%%%%
One important aspect of multi-probe analyses is the ability to offset systematic uncertainties, especially from astrophysics, that heavily impact individual probes. As an example, we consider intrinsic alignment of source galaxies, which has been studied in observations, simulations, and theory \citep[e.g.,][]{his04,mhi06, job10, smm14, tri14,tsm15, bvs15}. We extend the work presented in \cite{keb16} to include the effect of IA on galaxy-galaxy lensing
\be
\label{eq:CgI}
C_{\delta_{\mathrm{g}}\kappa}^{ij}(l) \rightarrow C_{\delta_{\mathrm{g}}\kappa}^{ij}(l) + C_{\delta_{\mathrm{g}}I}^{ij}(l) \,,
\ee
where
\be
C_{\delta_{\mathrm{g}}I}^{ij}(l) = -\int d\chi  \frac{q_{\delta_{\mathrm{g}}}^i}{\chi^2} \frac{n_{\mathrm{source}}^j(z)}{\bar{n}_{\mathrm{source}}^j}\frac{dz}{d\chi} b_\mathrm{g}^i(z) f_{\mathrm{red}}(z, m_{\mathrm{lim}})P_{\delta I}(l/\chi,z,m_{\mathrm{lim}}) \,,
\ee
with $z = z(\chi)$. The $j$ dependent term is the normalized distribution of source galaxies in redshift bin $j$, $f_{\mathrm{red}}$ is the fraction of red galaxies which is evaluated as a function of limiting magnitude $m_{\mathrm{lim}} = 27$, and $P_{\delta I}$ the cross power spectrum between intrinsic galaxy orientation and matter density contrast. 

The IA contamination of our data vector assumes a DEEP2 luminosity function \citep{fww07} and the tidal alignment scenario described in \cite{bvs15, keb16}. The tidal alignment scenario is in good agreement with observations; using the DEEP2 luminosity function should be considered as an upper limit  of the strength of IA contaminations. 

\begin{figure*}
  \includegraphics[width=8.5cm]{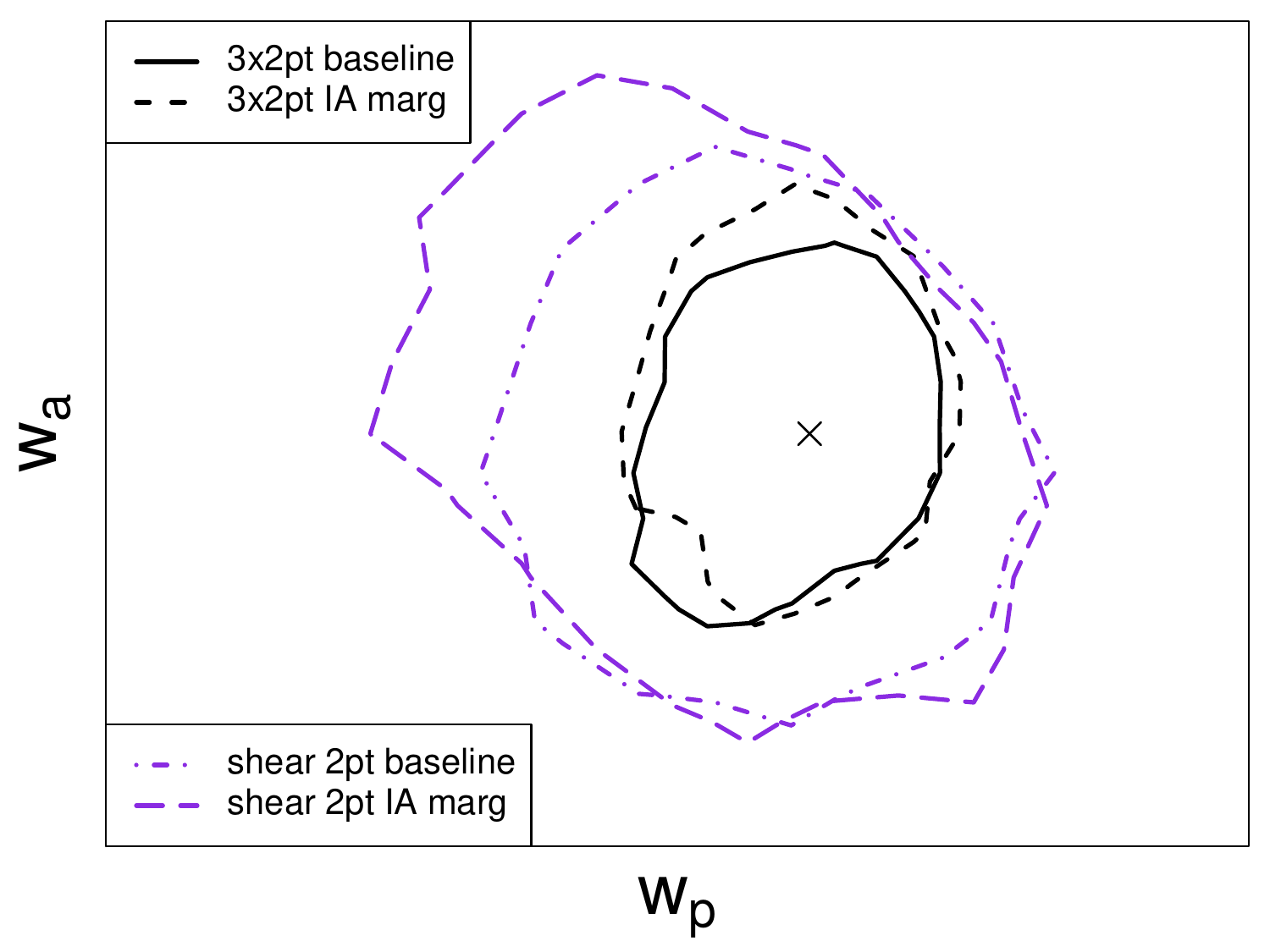}
   \includegraphics[width=8.5cm]{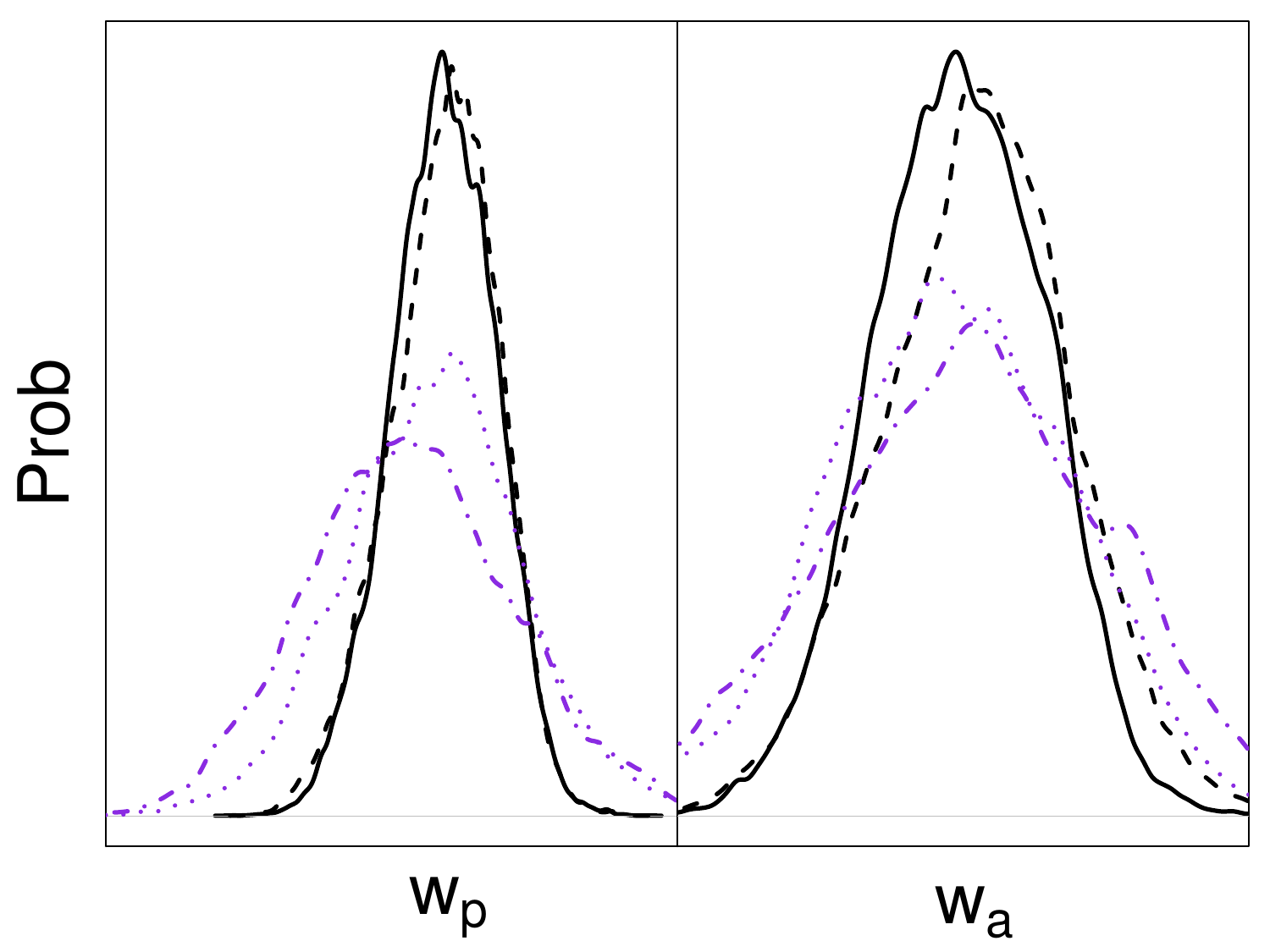}
\caption{We extend the baseline analysis of cosmic shear and 3x2pt \ti{(violett/dot-dashed and black/solid)} to IA mitigation scenarios \ti{(violett/long-dashed and black/dashed)}.}
 \label{fi:IA}
\end{figure*}

In Fig. \ref{fi:IA} we compare the baseline analysis for cosmic shear and 3x2pt (no IA contamination) to the case where IA contaminates the data vectors. In the latter case we marginalize over 10 nuisance parameters \citep[4 for IA and 6 for luminosity function uncertainties, see][for details]{keb16} to account for the IA contamination. Although we assume the tidal alignment scenario as a contaminant, we choose a different IA model for the marginalization (non-linear alignment with the \textsc{Halofit} fitting formula) to mimic a realistic analysis.   

We find that in the presence of multiple probes, photo-z, shear calibration and galaxy bias uncertainties, the assumption of an imperfect IA model in the marginalization is negligible. As expected when including 10 more dimensions in the analysis the constraints weaken but again the effect is not severe. Note that the 3x2pt data vector only includes galaxy-galaxy lensing tomography bins for which the photometric source redshifts are behind the lens galaxy redshift bin. Hence only a small fraction of source galaxies in the low-z tail of the redshift distribution contribute an IA signal to galaxy-galaxy lensing. As a consequence the 3x2pt data vector contains only marginally more information on IA, and improvements in the self-calibration of IA parameters is largely due to the enhanced constraining power on parameters which are degenerate with IA.

%%%%%%%%%%%%%%%%%%%%%%%%%%%%%%%%%%%
%%%%%%%%%%%%%%%%%%%%%%%%%%%%%%%%%%%
\section{Discussion}
\label{sec:discussion}
%%%%%%%%%%%%%%%%%%%%%%%%%%%%%%%%%%%
%%%%%%%%%%%%%%%%%%%%%%%%%%%%%%%%%%%
The first step in designing a multi-probe likelihood analysis is to specify the exact details of the data vector. This is far from trivial; the optimal data vector is subject to various considerations. 

\begin{itemize}
\item \tbf{Science case}  This paper focusses on time-dependent dark energy as a science case with the fiducial model being $\Lambda$CDM. If there was indication for time-dependence, the data vector can be optimized (tomography bins, galaxy samples, scales) such that it is most sensitive to these signatures. The same holds when extending the science case to e.g., neutrino physics and cosmological tests of gravity \citep{jjk15,baa15}.
\item \tbf{Prior information} Prior information (from external data sets) should only be included if individual analysis of these data sets yield compatible results. Tension between data sets indicates new physics or insufficient modeling of systematics and needs to be resolved before pursuing a joint analysis. It must also be considered whether the external data set is independent or correlated; the latter case requires a joint analysis, whereas in the former case is it permitted to multiply the individual posterior probabilities. We note that in the era of precision cosmology prior information from the CMB can longer be assumed to be independent of low-z probes. Even correlations between SN1a and the probes considered in this paper should be examined, since both affected by magnification effects and offsets in the photometric calibration.
\item \tbf{Probes} In terms of basic signal-to-noise considerations, it may seem highly desirable to include as many probes as possible into the data vector. However, every probe is associated with systematic uncertainties and while some of these uncertainties (e.g., instrumental) can be modeled similarly across all probes, the modeling accuracy of astrophysical uncertainties may vastly differ. Adding probes with weakly/unconstrained astrophysical uncertainties, that translate into additional model parameters with weak/non-existing priors, is penalized in model comparison, e.g. when computing the Bayes factor. In such cases it can be favorable to exclude the corresponding probe.

\item \tbf{Summary statistics} N-point functions in Fourier and Real space are an established way to quantifying the information content of probes that trace the density field. Models for Fourier space summary statistics, such as power-, bi-, tri-spectra, are faster to evaluate as a function of cosmology and systematics. A direct reconstruction of the E-mode shear spectra in Fourier however suffers from leakage (mixing) of E- and B-modes due to finite survey size, masking, pixelization and binning \citep{smi06, bec13}. 

Real space measurements of e.g., two-, three-, four-point functions are much less sensitive to masking effects and can cleanly separate E- and B-modes in case of cosmic shear \citep{sek10,eif11, bec13}. However, these methods are slower and require a precise computational implementation when Fourier-transforming the modeled spectra and, in particular, when Fourier-transforming corresponding covariances. 

\item \tbf{Minimum and maximum angular scales} The small-scale (high-$l$) limit of the data vector is largely determined by our ability to model baryonic effects, non-linear evolution of the matter density field and galaxy biasing. It is hence directly related to resolution and physics modeling requirements of numerical simulations and their post-processing. On large scales instrumental effects (camera field of view, chip gaps) can play an important role. It is critical to carefully weigh the gain in information compared to the required systematics modeling when pushing either of these boundaries.
  
\item \tbf{Number of redshift bins} Decision drivers are the accuracy of photometric redshifts, astrophysical systematics that vary strongly with redshift (e.g., galaxy bias for a galaxy sample with a complicated selection function, and galaxy intrinsic alignment), and the redshift sensitivity of the science case. Strong redshift dependence of systematics favors multiple narrow tomographic bins for self-calibration concepts; the same is true for dark energy models that vary strongly as a functions of redshift.    

\item \tbf{Number of angular bins} Similar to the number of redshift bins one needs to determine the threshold when information saturates for a given range in scales. This is of particular importance since recent studies find strong correlation across bins, even in Fourier space, \citep[e.g.,][]{sht09}. The number of total bins in the joint data vector can also be limited by the number of independent realizations that can be generated for covariance estimation \citep{dos13,tjk13,hss07}. We recommend exploring methods for multi-probe data compression \citep[e.g.,][]{eks14} and advanced estimation concepts for covariances \citep{pos08}.  
\end{itemize}

%%%%%%%%%%%%%%%%%%%%%%%%%%%%%%%%%%%%%%%%%%%%%%%%%%
\section{Conclusions}
\label{sec:conc}
%%%%%%%%%%%%%%%%%%%%%%%%%%%%%%%%%%%%%%%%%%%%%%%%%%
The joint and consistent modeling of cosmological probes including their correlated signals and systematics is one of the main challenges for ongoing and (even more) for future surveys. In this paper we present simulated likelihood analyses for an LSST like data set that include cosmological information from cosmic shear, galaxy-galaxy lensing, galaxy clustering (including high-z photometric BAOs), cluster number counts and cluster weak lensing. We also include a variety of systematic effects that degrade the cosmological constraining power such as uncertainties in photo-z calibration (for source and lens bins), shear calibration, cluster mass-observable relation, galaxy bias, and galaxy intrinsic alignment.

Although our simulated constraints are dominated by systematic uncertainties (see Fig. \ref{fi:galvary}), we stress that the joint analysis of multiple probes nonlinearly increases the cosmological information when compared to single probe analysis (see Fig. \ref{fi:moneyplot}). In order to further optimize the multi-probe ansatz, we consider several variations of our baseline analysis: 

We find that small-scale clustering information adds valuable information, and if sufficiently accurate models can be developed, it should be included either by including non-linear scales in the clustering 2-point function or by including galaxy cluster weak lensing in the analysis (Fig. \ref{fi:HOD}). Increasing the number density of galaxies however is not the most promising way to increase cosmological information from photometric surveys (Fig. \ref{fi:galvary}), in particular since this avenue of survey optimization goes along with increased systematics. In contrast a smaller, but well characterized, sample of galaxies that is spread out over a large area appears favorable. 

It is not the intent of this paper to present a complete analysis of these effects, but to demonstrate the ability of the newly developed \textsc{CosmoLike} software to model such complex analyses. The fiducial values for systematic uncertainties described in Table \ref{tab:params} are currently not achievable, but assume substantial improvements in e.g. photo-z and shear estimation algorithms by the end of the LSST survey. As updated values for Table \ref{tab:params} become available, \textsc{CosmoLike} can be used to evaluate the improvements or degradation in constraining power for some of the most complex multi-probe analyses; we recommend interfacing \textsc{CosmoLike} with the \textsc{CosmoSIS} framework \citep{zpj15} to further increase modeling options. We emphasize that the impact of systematics should be examined with respect to the most stringent statistical uncertainties, which will come from a joint analysis, and in the context of other systematics that are present. Realistic modeling of these scenarios allows us to define realistic requirements on future surveys. 

Improving the parameterization and parameter priors of systematics is a priority to maximize cosmological information. Constraints on exciting physics are only possible if systematics are well understood. In this context it is interesting to examine whether the information gained from Null-tests can be included a priori in the data vector or the modeling framework. For example, a non-zero clustering signal in cross-tomographic bins due photo-z errors  can be used as `data'. The amplitude of this signal depends (obviously) on the accuracy of photo-z's but it also depends on the amplitude of the clustering signal. Cross-tomographic clustering can hence be included as a signal that is modeled with a parameter describing the `leakage fraction of galaxies'. Results from null-tests that do not depend on cosmology (e.g. star-galaxy correlation functions), can be incorporated into the analysis as priors on systematics. 

The tightest constraints on the physics of the Universe will be obtained by combining multiple probes of the Large Scale Structure of the Universe that differ in terms of underlying physics and are affected by different systematics. As we approach the regime of systematics limited surveys every source of information is valuable. The inclusion of various CMB probes, SN1a, Strong Lensing, spectroscopic BAOs and Redshift Space Distortions, void and trough lensing \citep{kcd13,mss14,gfa16}, present multiple opportunities to increase the data vector considered here (not to mention higher-order summary statistics).

\section*{Acknowledgments}
We thank Sarah Bridle, Scott Dodelson, Joe Zuntz, Risa Wechsler, Gary Bernstein, Bhuvnesh Jain, and Josh Frieman for frequent and fruitful discussions on the topic of multi-probe cosmology. EK thanks Benjamin Joachimi and Joe Zuntz and for code-comparison efforts. EK acknowledges support from NSF grant AST-0908027. This paper is based upon work supported in part by the National Science Foundation under Grant No. 1066293 and the hospitality of the Aspen Center for Physics. Part of the research was carried out at the Jet Propulsion Laboratory (JPL), California Institute of Technology, under a contract with the National Aeronautics and Space Administration. All computations were performed on the JPL High Performance Computing systems; we thank the JPL Super Computing Consult team for outstanding support.

%%%%%%%%%%%%%%%%%%%%%%%%%%%%%%%%%%%%%%%%%%%%%%%%%%



\bibliographystyle{mnras}
\bibliography{multi-probe-paper}

%%%%%%%%%%%%%%%%%%%%%%%%%%%%%%%%%%%%%%%%%%%%%%%%%%

%%%%%%%%%%%%%%%%% APPENDICES %%%%%%%%%%%%%%%%%%%%%
\begin{appendix}
\section{Covariance implementation}
\label{sec:app1}
%%%%%%%%%%%%%%%%%%%%%%%%%%%%%%%%%%%%%%%%%%%%%%%%%%%%%%%%%%%%%%%%%%%%%%%%%%%%%%

\renewcommand{\arraystretch}{1.5}
\begin{table*}
\label{tab:halo}
\caption{Redshift weight functions, halo model building blocks, and noise terms for different probes.}
\begin{tabular*}{2\columnwidth}{clcccll}
\hline
\hline
observable $A$ & model & weight $q_A^i$ & large-scale bias $b_{A}^i$ & $\tilde{u}_{A}^i(k,M)$ & power spectrum $P_{AB}$ & noise $N_A^i$\\
\hline
 $\kappa$ &  & Eq.~(\ref{eq:qkappa})  & 1 & $\frac{M}{\bar{\rho}}\tilde{u}_{m}(k,M)$ & $P_{\kappa B} = P_{\mathrm{m} B}$ & $\sigma_\epsilon^2/\bar{n}_{\mathrm{source}}^i$\\
$\delta_{\lambda_\alpha}$& & Eq.~(\ref{eq:qlambda})& $ \frac{ \int dM  \frac{dn}{dM} b_{\mathrm{h}}(M)\int_{\lambda_{\alpha,\mathrm{min}}}^{\lambda_{\alpha,\mathrm{max}}} d\lambda \,p(M | \lambda,z)}{\int dM   \frac{dn}{dM}\int_{\lambda_{\alpha,\mathrm{min}}}^{\lambda_{\alpha,\mathrm{max}}} d\lambda p(M | \lambda,z)}$& $  \frac{\int_{\lambda_{\alpha,\mathrm{min}}}^{\lambda_{\alpha,\mathrm{max}}} d\lambda \,p(M | \lambda,z)}{\int dM   \frac{dn}{dM}\int_{\lambda_{\alpha,\mathrm{min}}}^{\lambda_{\alpha,\mathrm{max}}} d\lambda p(M | \lambda,z)}$ &Eq.~(\ref{eq:Pcluster}) &$\Omega_{\mathrm{s}}/\mathcal{N}^i(\lambda_\alpha) $\\
$\delta_\mathrm{g}$& linear bias &Eq.~(\ref{eq:qg}) & $b_\mathrm{g}^i$ &$b_\mathrm{g}^i\frac{M}{\bar{\rho}}\tilde{u}_{m}(k,M)$ & $P_{\mathrm{g} B} = b_\mathrm{g}^i P_{\mathrm{m} B}$ & $1/\bar{n}_{\mathrm{lens}}^i$\\
$\delta_\mathrm{g}$ & HOD $N_{\mathrm{g}}(M)$& Eq.~(\ref{eq:qg})& $\frac{\int dM \frac{dn}{dM} b(M) N_{\mathrm{g}}(M)}{\int dM \frac{dn}{dM} N_{\mathrm{g}}(M)} $ &$\frac{N_{\mathrm{c}}(M)\left[f_{\mathrm{c}} + N_{\mathrm{s}}(M)\tilde{u}_{\mathrm{s}}(k,M)\right)]}{\int dM \frac{dn}{dM}\ensav{N_\mathrm{g}(M)}}
$ & Eq.~(\ref{eq:PHOD}) & $1/\bar{n}_{\mathrm{lens}}^i$\\
\hline
\end{tabular*}
\end{table*}
\renewcommand{\arraystretch}{1.0}
The multi-probe covariance presented in Fig.~\ref{fi:covstruct} generalizes the non-Gaussian covariance terms previously described for cosmic shear \citep[e.g.][]{coh01,sht09,tah13}, and the joint analysis of cosmic shear and cluster counts \citep{tas14,sts14} to the set of LSS tracers considered in this paper.
We calculate the covariance of two angular power spectra as the sum of the Gaussian covariance, $ \mr{Cov^G}\left( C(l_1), C(l_2) \right)$, and non-Gaussian covariance in the absence of survey window effects, $\mr{Cov^{NG,0}}\left( C(l_1), C(l_2) \right)$, and the super-sample covariance, $\mr{Cov^{SSC}}\left( C(l_1), C(l_2) \right)$, which describes the uncertainty induced by large-scale density modes outside the survey window:
\begin{widetext}
\be
\mr{Cov}\left( C (l_1), C(l_2) \right)  = \mr{Cov^G}\left( C(l_1), C(l_2) \right)+\mr{Cov^{NG,0}}\left( C(l_1), C(l_2) \right)+\mr{Cov^{SSC}}\left( C(l_1), C(l_2) \right)\,.
\ee

The Gaussian covariance of two multi-probe power spectra is given by \citet{huj04}
\be
\mr{Cov^G}\left( C_{AB}^{ij} (l_1), C_{CD}^{kl} (l_2) \right) = \frac{4 \pi \delta_{l_1 l_2}}{ \Omega_{\rm{s}} (2l_1+1) \Delta l_1}  
\left[\left(C_{AC}^{ik}(l_1)+ \delta_{ik}\delta_{AC} N_{A}^i\right) \left(C_{BD}^{jl}(l_2)+ \delta_{jl}\delta_{BD}N_{B}^j\right) 
+\left(C_{AD}^{il}(l_1)+ \delta_{il}\delta_{AD} N_{A}^i\right) \left(C_{BC}^{jk}(l_2)+ \delta_{jk}\delta_{BC} N_{B}^j\right) \right]\,,
\ee
with the probe-specific noise terms $N_A^i$ given in Table~\ref{tab:halo}.

The non-Gaussian covariance in the absence of survey window effects is calculated as the projected trispectrum, 
\be
\mr{Cov^{NG,0}}\left( C_{AB}^{ij} (l_1), C_{CD}^{kl} (l_2) \right)  = \frac{1}{\Omega_{\mr s}} \int_{|\mathbf l|\in l_1}\frac{d^2\mathbf l}{A(l_1)}\int_{|\mathbf l'|\in l_2}\frac{d^2\mathbf l'}{A(l_2)} 
 \int \d \chi \;\frac{q_{A}^i(\chi) q_{B}^j(\chi) q_{C}^k(\chi) q_{D}^l(\chi)}{\chi^6}\;\;T_{ABCD}^{ijkl}\left(\mathbf l/\chi,-\mathbf l/\chi,\mathbf l'/\chi,-\mathbf l'/\chi; z(\chi)\right)
\ee
where we approximate the $ABCD$ trispectrum as the sum of the linearly biased $(2+3+4)$-halo matter trispectrum \citep[see e.g.][for details]{cos02, taj09}, and a probe-specific 1-halo trispectrum:
\be
T_{ABCD}^{ijkl}\left(\mathbf k,-\mathbf k,\mathbf k',-\mathbf k';z\right) \approx  b_{A}^i(z) b_{B}^j(z) b_{C}^k(z) b_{D}(z)T_{\mathrm{m}}^{4\mathrm{h}+3\mathrm{h}+2\mathrm{h}}\left(\mathbf k,-\mathbf k,\mathbf k',-\mathbf k'; z\right)+ T_{ABCD}^{ijkl,1\mathrm{h}}\left(k,k,k',k'; z\right) 
\ee
where we introduced $b_\kappa  = 1$ for convenience, and the one-halo trispectrum $ T_{ABCD}^{ijkl,1\mathrm{h}}$:
\begin{eqnarray} 
\label{eq:T1h} \#_{\delta_\lambda}  \le 1: & T_{ABCD}^{ijkl,1\mathrm{h}}(k,k,k',k';z) &= \int dM \frac{dn}{dM} \ensav{\tilde{u}_{A}^i(k,M)\tilde{u}_{B}^j(k,M) \tilde{u}_{C}^k(k',M)\tilde{u}_{D}^l(k',M)}\\
\label{eq:T1cc} \#_{\delta_\lambda}  =2 : & T_{\delta_{\lambda_{\alpha}}B\delta_{\lambda_{\beta}}D}^{ijkl,1\mathrm{h}}(k,k,k',k';z) &=\delta_{\alpha,\beta}\int dM \frac{dn}{dM} \tilde{u}_{\delta_{\lambda_{\alpha}}}(k,M)\tilde{u}_{B}^j(k,M) \tilde{u}_{\delta_{\lambda_{\beta}}}(k',M)\tilde{u}_{D}^l(k',M)\,,
\end{eqnarray}
using the observable specific halo model building blocks given in Table~\ref{tab:halo}; $\#_{\delta_\lambda}$ is the multiplicity of the cluster density contrast in $\{ABCD\}$, and the special case in Eq.~(\ref{eq:T1cc}) enforces the vanishing of the one-halo term between two different clusters. The ensemble average in Eq.~(\ref{eq:T1h}) only comes into effect on moments of the HOD, which we evaluate assuming that satellite galaxies are Poisson distributed.

The super-sample covariance describes the response of the summary statistics to a large scale background density mode; adapting the notation of \citet{tah13,sts14} to the multi-probe power spectrum case, it is given by
\be
\mr{Cov^{SSC}}\left( C_{AB}^{ij} (l_1), C_{CD}^{kl} (l_2) \right) = \int \d \chi \;\frac{q_{A}^i(\chi) q_{B}^j(\chi) q_{C}^k(\chi) q_{D}^l(\chi)}{\chi^4}\;\;\frac{\partial P_{AB}(l_1/\chi,z(\chi))}{\partial \delta_{b}}\frac{\partial P_{CD}(l_2/\chi,z(\chi))}{\partial \delta_{b}}\sigma_b(\Omega_{\rm{s}};z(\chi))\,,
\ee
with $\sigma_b(\Omega_{\rm{s}},z(\chi))$ the variance of the background mode over the survey window, 
\be
\sigma_b(\Omega_{\rm{s}};z) = \int \frac{d^2k_\perp}{(2\pi)^2} P_{\mathrm{lin}}(k_\perp,z)|\tilde{W}_{\mathrm{s}}(k_\perp,z)|^2 \approx\int \frac{d^2k_\perp}{(2\pi)^2} P_{\mathrm{lin}}(k_\perp,z)\left[\frac{2J_1(k_\perp\chi(z)\theta_\mathrm{s})}{k_\perp\chi(z)\theta_\mathrm{s}}\right]^2\,,
\ee
where in the second step we approximated the survey window function $W_\mathrm{s}$ assuming a disk-like survey geometry of radius $\theta_{\mathrm{s}} = \sqrt{\Omega_{\mathrm{s}}/\pi}$.
In order to evaluate the response of the multi-probe power spectra using the halo model and peak background split analogously to the matter power spectrum derivation \citep{tah13, lht14, cws14}, we rewrite the usual auxiliary halo model function $I^\alpha_\beta(k_1,...,k_\beta)$ for the multi-probe case, noting that the power spectrum only requires $\beta\le 2$:
\bea
\label{eq:IAB}
I^\alpha_A(k) = \int dM \frac{dn}{dM} b_{\mathrm{h},\alpha}(M)\ensav{\tilde{u}_{A}^i(k,M)}\;, && I^\alpha_{AB}(k,k') = \int dM \frac{dn}{dM} b_{\mathrm{h},\alpha}(M)\ensav{\tilde{u}_{A}^i(k,M)\tilde{u}_{B}^i(k',M)}\,,
\eea
where $b_{\mathrm{h},\alpha}$ is the $\alpha$-th order halo bias, with $b_{\mathrm{h},0} =1$, and where we again neglect higher-order biasing, i.e. $b_{\mathrm{h},\ge2} =0$. In this notation, the halo model description of the multi-probe power spectra is given by
\be
P_{AB}(k,z) =P^{2\mathrm{h}}_{AB}(k,z)+P^{1\mathrm{h}}_{AB}(k,z)=  P_{\mathrm{lin}}(k,z) I^1_A(k) I^1_B(k) + I^0_{AB}(k,k)\,.
\ee
The response of the matter power spectrum is given by \citep{tah13, lht14, cws14}
\be
\frac{\partial P_{\mathrm{mm}}(k;z)}{\partial \delta_{b}} = \left(\frac{68}{21}-\frac{1}{2}\frac{d\ln k^3\,P_{\mathrm{lin}}(k,z)}{d \ln k} \right) I^1_\mathrm{m}(k) I^1_\mathrm{m}(k)P_{\mathrm{lin}}(k,z)+ I^1_{\mathrm{mm}}(k,k).
\ee
To calculate the response of galaxy-galaxy lensing and clustering power spectra, one needs to account for the fact that the galaxy density contrast is estimated using the mean galaxy density within the survey window, and the observed power spectrum $\hat{P}_{\delta_\mathrm{g}B}(k)$ is rescaled with respect to the cosmic mean, $\hat{P}_{\delta_\mathrm{g}B}(k) = P_{\delta_\mathrm{g}B}(k)/(1+b_\mathrm{g})^n$, with $n = 2$ if $B = \delta_\mathrm{g}$ and $n = 1$ otherwise. For large surveys, the effect of this rescaling of the power spectrum is negligible, but does affect the response to large-scale modes:
\be
\frac{\partial \hat{P}_{\delta_\mathrm{g}B}(k)}{\partial \delta_{b}} \approx \frac{\partial P_{\delta_\mathrm{g}B}(k)}{\partial \delta_{b}} - n b_\mathrm{g}P_{\delta_\mathrm{g}B}(k)\,.
\ee
For the HOD description of galaxy biasing, and the halo model description of cluster lensing, this rescaling occurs automatically as the radial profile functions $\tilde{u}(k)$ in Eq.~(\ref{eq:IAB}) are already weighted by the local mean of these objects (if the HOD, or the cluster mass-richness relation are estimated from the same survey). Applying the same peak background split calculation to the (implicit) denominator of Eq.~(\ref{eq:IAB}) as well, we arrive at
\be
\frac{\partial P_{AB}(k,z)}{\partial \delta_{b}} = \left(\frac{68}{21}-\frac{1}{2}\frac{d\ln k^3\,P_{\mathrm{lin}}(k,z)}{d \ln k} \right) I^1_A(k) I^1_B(k)P_{\mathrm{lin}}(k,z)+ I^1_{AB}(k,k)-\Big[b_{A,A\ne\kappa} +b_{B,B\ne\kappa}\Big]P_{AB}(k,z)\,.
\ee

The covariance of two cluster number count bins is given by the sum of a shot noise, and a super-sample variance term,
\be
\mathrm{Cov}\left(\mathcal{N}^i_{\lambda_\alpha},\mathcal{N}^j_{\lambda_\beta}\right) = \delta_{i,j} \delta_{\alpha,\beta} \mathcal{N}^i_{\lambda_\alpha} + \Omega_{\mathrm{s}}^2 \,\int d \chi q_{\lambda_\alpha}^i(\chi) q_{\lambda_\beta}^j(\chi)\left[\int dM \frac{dn}{dM} b_{\mathrm{h}}(M,z)\int_{\lambda_{\alpha,\mathrm{min}}}^{\lambda_{\alpha,\mathrm{max}}} d\lambda\, p(M | \lambda,z)\right] \left[\int dM' \frac{dn}{dM'} b_{\mathrm{h}}(M',z)\int_{\lambda_{\beta,\mathrm{min}}}^{\lambda_{\beta,\mathrm{max}}} d\lambda'\, p(M' | \lambda',z)\right]\,,
\ee
where we have neglected correlations across redshift bins.

We approximate the covariance between cluster number counts and multi-probe power spectra by the dominant super-sample contribution \citep[but see][for a discussion of other terms]{tas14,sts14}
\be
\mathrm{Cov}\left(\mathcal{N}^i_{\lambda_\alpha},C_{AB}^{jk}(l)\right) = \Omega_\mathrm{s}\; \int d\chi \frac{q_{\lambda_\alpha}^i(\chi)q_A^j(\chi) q_B^k(\chi)}{\chi^2 }\left[\int dM \frac{dn}{dM} b_{\mathrm{h}}(M,z)\int_{\lambda_{\alpha,\mathrm{min}}}^{\lambda_{\alpha,\mathrm{max}}} d\lambda\, p(M | \lambda,z)\right] \frac{\partial P_{AB}(k,z(\chi))}{\partial \delta_{b}} \sigma_b(\Omega_{\rm{s}};z(\chi))\,.
\ee
\end{widetext}
%with $n^i$ the number of source galaxies in tomography bin $i$, $\sigma_\epsilon$ the ellipticity dispersion, 
%$A(l_i) = \int_{|\mathbf l|\in l_i}d^2\mathbf l \approx 2 \pi l_i\Delta l_i$ the integration area associated with a 
%power spectrum bin centered at $l_i$ and width $\Delta l_i$, and $T_{\kappa,0}$ and $T_{\kappa,\rm{HSV}}$ 
%the convergence trispectrum in the absence of finite volume effects and the halo sample variance contribution 
%to the trispectrum \citep{sht09,takada2013}. Our halo model implementation for these terms is described in \citet{ekd14}.
\end{appendix}

%%%%%%%%%%%%%%%%%%%%%%%%%%%%%%%%%%%%%%%%%%%%%%%%%%

% Don't change these lines
\bsp	% typesetting comment
\label{lastpage}
\end{document}